\documentclass{article} 
\usepackage{iclr2027_conference,times}

\usepackage{amsmath,amsfonts,bm}

\def\eqref#1{equation~\ref{#1}}
\def\Eqref#1{Equation~\ref{#1}}

\def\1{\bm{1}}

\DeclareMathAlphabet{\mathsfit}{\encodingdefault}{\sfdefault}{m}{sl}
\SetMathAlphabet{\mathsfit}{bold}{\encodingdefault}{\sfdefault}{bx}{n}

\usepackage{booktabs}   
\usepackage{array}      
\usepackage{multirow}   
\usepackage{longtable}  
\usepackage{wrapfig}
\usepackage{graphicx}
\usepackage{subcaption}
\usepackage{placeins}
\usepackage{float}
\usepackage{microtype}
\usepackage{url}
\usepackage{hyperref}

\title{EGGPlacer: Energy-Guided Generative Chip Placement with Diffusion Models}

\author{
	\textbf{Qi Zhou, Guojun Liu\thanks{Corresponding author.}, Juanhong Zhao, Guangzhi Qi, Zehao Chen, Ming Lu}\\
	Faculty of Computing, Harbin Institute of Technology
}

\hypersetup{
    pdftitle={EGGPlacer: Energy-Guided Generative Chip Placement with Diffusion Models},
    pdfauthor={Qi Zhou, Guojun Liu, Juanhong Zhao, Guangzhi Qi, Zehao Chen, Ming Lu}
}

\iclrfinalcopy
\begin{document}

\maketitle
\lhead{Preprint}

\begin{abstract}
Chip placement strongly affects routed wirelength and thermal behavior, yet these downstream metrics are difficult to evaluate during macro placement.
Learning-based macro placers commonly optimize the macro-only half-perimeter wirelength (HPWL), denoted mHPWL, which ignores standard cell connectivity and spatial demand. In contrast, analytical mixed-size placers optimize differentiable full netlist wirelength and density objectives but are typically unaware of thermal effects.
We propose EGGPlacer, an energy-guided diffusion framework that incorporates analytical placement feedback and thermal feedback into macro placement generation.
An analytical placer acts as an online physical response model, exposing the standard cell distribution and a full netlist wirelength proxy induced by each intermediate macro placement, while a differentiable thermal model provides temperature feedback.
These signals are formulated as composable energies and injected throughout reverse diffusion without retraining the underlying generative model.
Experiments on public placement benchmarks show that EGGPlacer achieves competitive mixed-size HPWL relative to strong analytical and learning-based macro placers. In the end-to-end evaluation, EGGPlacer also obtains the lowest average routed wirelength and peak temperature among the evaluated methods.
Ablation studies further demonstrate that analytical and thermal guidance provide distinct and complementary benefits across wirelength proxies, routed wirelength, and thermal objectives.
\end{abstract}

\section{Introduction}
\label{Introduction}

\begin{wrapfigure}{r}{0.44\linewidth}
	\vspace{-0.8em}
	\centering
	\includegraphics[width=\linewidth]{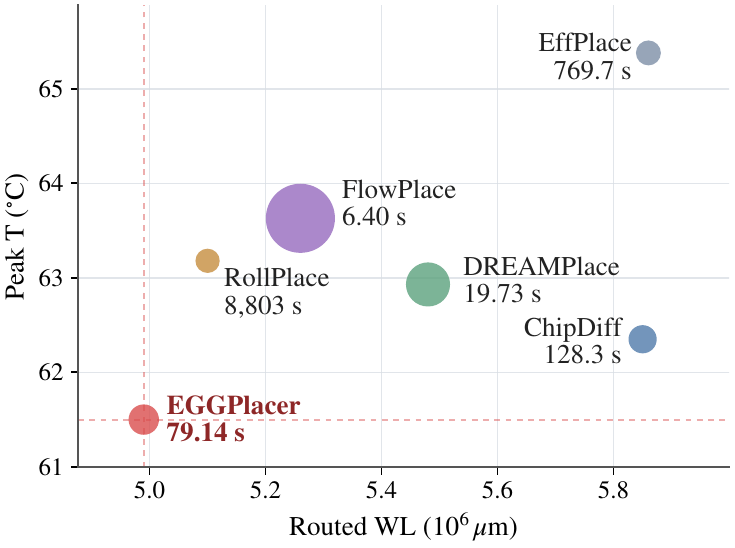}
	\caption{Routed wirelength (Routed WL), peak temperature (Peak T), and runtime for EGGPlacer and the evaluated baselines. Lower values on both axes are better, and larger circles indicate lower runtime.}
	\label{fig:quality-thermal-efficiency-tradeoff}
\end{wrapfigure}

Macro placement is a critical stage in chip physical design, where macros of heterogeneous sizes must be arranged on the chip canvas while satisfying geometric and physical constraints.
The resulting macro configuration strongly affects downstream design quality, including routed wirelength, timing, power, and thermal behavior.
For example, shorter routed wirelength reduces the interconnect resources required for routing, while controlling temperature helps limit leakage power and preserve chip reliability~\citep{skadron2004temperature,han2021hotspot7,srinivasan2004lifetime}.
However, accurately evaluating a macro placement is challenging because the final values of these downstream metrics depend on subsequent standard cell placement, legalization, and routing.
This coupling makes it difficult to determine whether a promising macro configuration will eventually lead to a high-quality physical design.
Figure~\ref{fig:quality-thermal-efficiency-tradeoff} previews the central empirical result: EGGPlacer achieves strong combined results in routed wirelength, peak temperature, and runtime.

\begin{figure}[!t]
	\centering
	\includegraphics[width=\linewidth]{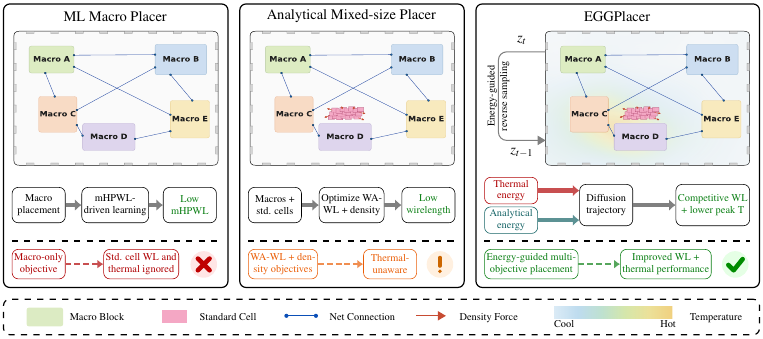}
	\caption{Motivation for EGGPlacer. Learning-based macro placers primarily
		optimize mHPWL and omit standard cell contributions and thermal
		behavior. Analytical mixed-size placers optimize a full netlist wirelength proxy
		and density but remain unaware of thermal effects. EGGPlacer injects analytical
		and thermal energies into reverse diffusion to optimize wirelength and
		temperature.}
	\label{fig:framework}
	\vspace{-2em}
\end{figure}

The challenge becomes increasingly pronounced for modern very large scale integration (VLSI) circuits containing hundreds of macros and a large number of standard cells.
The combinatorial search space grows rapidly with circuit scale, while repeatedly invoking downstream physical design tools to evaluate candidate macro placements is prohibitively expensive.
A practical macro generator therefore needs feedback that is both informative about downstream behavior and inexpensive enough to evaluate repeatedly while macro decisions are still being made.
To reduce the cost of exploring the large placement space, recent machine learning-based placers learn structural patterns from netlists and placement data to generate macro configurations efficiently.

Despite these advances, most existing learning-based macro placers are still optimized primarily using the macro-only variant of half-perimeter wirelength (HPWL), hereafter mHPWL~\citep{maskplace,chipformer,efficientplace,RollPlace,flowplace,chipdiffusion}.
This metric is attractive because it can be evaluated efficiently during macro placement, but it provides only an indirect proxy for downstream design quality.
For example, a placement with low mHPWL may leave insufficient or poorly distributed whitespace for standard cells, while spatially concentrated power dissipation can lead to thermal hotspots.
Analytical mixed-size placers address the first limitation by optimizing a differentiable full netlist wirelength proxy together with density after standard cells are introduced~\citep{nonlinear6,nonlinear7eplace,nonlinear8replace,dreamplaceDAC}. The resulting signal captures interactions between macros and standard cells that mHPWL omits. However, the conventional analytical objective remains unaware of temperature after placement. Thus, neither the learning pipeline driven by mHPWL nor the analytical pipeline driven by wirelength proxies and density exposes both standard-cell-aware placement feedback and thermal behavior to a learned macro generator.

Diffusion models~\citep{DDPM,DDIM} construct placements through iterative denoising.
ChipDiffusion uses guided sampling to generate macro placements for unseen circuits and represents standard cell clusters as virtual macros, while FlowPlace uses flow matching and an expert prior favoring macro placement near the layout boundary~\citep{chipdiffusion,flowplace}.
However, neither method explicitly models the downstream standard cell placement response or how thermal distributions vary with macro placement.
Their iterative trajectories nevertheless provide intervention points for injecting downstream physical feedback before the final configuration is determined.

To expose downstream standard-cell-aware physical and thermal feedback during macro generation, we propose \textbf{EGGPlacer}, an \textbf{E}nergy-\textbf{G}uided \textbf{G}enerative \textbf{Placer} based on diffusion models that incorporates multiple placement objectives into the reverse diffusion process through differentiable energy functions.
As illustrated in Figure~\ref{fig:framework}, these energy functions provide a unified interface for transferring downstream signals into macro generation. Analytical guidance exposes the standard cell response and a full netlist wirelength proxy, while thermal guidance exposes temperature changes caused by placement.
Applying both energies repeatedly during reverse sampling lets the learned placement prior perceive objectives that mHPWL alone cannot express, rather than waiting for a subsequent flow to reveal them.

Our main contributions are summarized as follows:

\begin{itemize}
	\item We identify a gap in objective visibility across learning-based macro placement driven by mHPWL, analytical placement driven by wirelength proxies and density, and downstream thermal evaluation, and bridge it with EGGPlacer, which exposes these complementary signals to macro generation through composable energy guidance.
	\item We introduce analytical and differentiable thermal guidance within a unified reverse sampling process. Analytical guidance transfers a full netlist wirelength proxy, density, and spatial competition between macros and standard cells, while thermal guidance exposes temperature changes caused by placement.
	\item We empirically show that the fidelity of HPWL as a routed wirelength proxy varies across placement engines, even with closely related analytical formulations. This finding highlights the importance of aligning macro placement generation with the downstream design flow.
\end{itemize}

\section{Problem Formulation}
\label{Problem Formulation}

Given a chip placement instance, we model the circuit as a hypergraph $H = (V, E)$. The subsets $V_M$, $V_S$, and $V_F$, whose union is $V$, represent movable macros, standard cells, and fixed I/O pins, respectively, while $E$ denotes the set of hyperedges interconnecting these vertices. Each movable cell $v_i$ is assigned a fixed width $w_i$ and height $h_i$, and all cells must be positioned within a prescribed 2D placement region $\Omega$. This work targets macro placement generation, namely predicting the horizontal and vertical center coordinate vectors $\boldsymbol{x}^M$ and $\boldsymbol{y}^M$ for all movable macros.

We write the general macro placement problem as
\begin{equation}
	\begin{aligned}
		\min_{\boldsymbol{x}^{M},\boldsymbol{y}^{M}} \quad & \mathcal{J}_{M}\big(H;\boldsymbol{x}^{M},\boldsymbol{y}^{M}\big) \\
		\text{s.t.} \quad & \operatorname{Overlap}\big(\boldsymbol{x}^{M},\boldsymbol{y}^{M},\boldsymbol{w}^{M},\boldsymbol{h}^{M}\big) = 0, \\
		& \operatorname{Boundary}\big(\boldsymbol{x}^{M},\boldsymbol{y}^{M},\boldsymbol{w}^{M},\boldsymbol{h}^{M},\Omega\big) = 0,
	\end{aligned}
	\label{eq:macro_placement}
\end{equation}

\begin{wrapfigure}{r}{0.41\linewidth}
	\centering
	\includegraphics[width=0.99\linewidth]{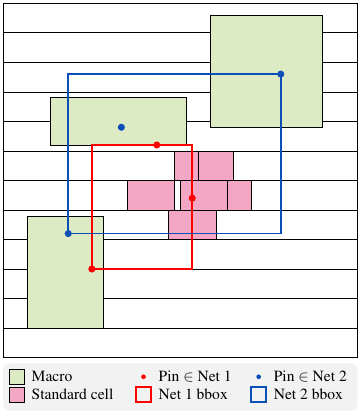}
	\caption{An example chip placement. Green blocks denote macros, pink
		blocks denote standard cells, and colored boxes bound the pins included for
		each net.}
	\label{fig:PlacementExample}

\end{wrapfigure}

where $\mathcal{J}_{M}$ denotes a generic macro placement objective.
The nonoverlap constraint enforces zero spatial overlap between any pair of macros, whereas the boundary constraint ensures all macros reside within the feasible placement region $\Omega$.
mHPWL is commonly used as the objective $\mathcal{J}_{M}$ in learning-based macro placement, i.e., $\mathcal{J}_{M}=\operatorname{mHPWL}$~\citep{maskplace,chipformer,efficientplace,wiremaskbbo,RollPlace}.
As illustrated in Figure~\ref{fig:PlacementExample}, the HPWL of a net is half the perimeter of the smallest axis-aligned bounding box enclosing the pins included in the calculation.
mHPWL forms this box using only pins on macros and fixed terminals, whereas full netlist HPWL also includes pins on standard cells after they are placed.
Both are placement-stage wirelength proxies computed without routing.
Net~1 illustrates a standard cell pin that contributes to full netlist HPWL but is absent from mHPWL, while the illustrated pins of Net~2 belong to macros.
Since standard cells typically far outnumber macros, ignoring their connectivity and spatial demand can make mHPWL a poor indicator of the resulting mixed-size placement.

Analytical global placement formulates mixed-size placement as continuous optimization over the coordinates of movable macros and standard cells~\citep{nonlinear7eplace,nonlinear8replace,dreamplaceDAC}. Let $\boldsymbol{x}=[\boldsymbol{x}^{M};\boldsymbol{x}^{S}]$ and $\boldsymbol{y}=[\boldsymbol{y}^{M};\boldsymbol{y}^{S}]$ denote the coordinate vectors of all movable cells. A typical objective combines differentiable wirelength and density terms:
\begin{equation}
	\mathcal{J}_{\mathrm{ana}}(\boldsymbol{x},\boldsymbol{y})
	=
	\sum_{e\in E} \widetilde{\operatorname{WL}}\big(e;\boldsymbol{x},\boldsymbol{y}\big)
	+ \lambda_D \, D(\boldsymbol{x},\boldsymbol{y}),
	\label{eq:global_placement}
\end{equation}
where $\widetilde{\operatorname{WL}}(e;\boldsymbol{x},\boldsymbol{y})$ denotes a differentiable wirelength proxy for net $e$, $D$ is the density penalty, and $\lambda_D$ controls the tradeoff between wirelength and density. This objective captures full netlist connectivity together with spatial competition among movable cells.

The distinction in Figure~\ref{fig:framework} defines the problem addressed by EGGPlacer. \Eqref{eq:macro_placement} is the general macro placement formulation, mHPWL is the objective commonly used by learning-based placers, and \Eqref{eq:global_placement} exposes a differentiable full netlist wirelength proxy and standard cell density. EGGPlacer therefore seeks macro placements that reduce full netlist wirelength and thermal cost while accounting for macro feasibility and standard cell spatial demand. Rather than treating these terms as a standalone placement objective, we formulate them as guidance energies and inject the resulting corrections into the reverse diffusion trajectory.

\section{Method}
\label{Method}

\subsection{Energy-Guided Diffusion Placement}
\label{sec:energy_guided_diffusion}

Let $\mathbf{z}_0\in\mathbb{R}^{N_M\times 2}$ denote the normalized coordinates of $N_M$ movable macros, and let $\mathbf{c}$ denote the netlist and physical design context. We use a standard noise prediction diffusion model~\citep{DDPM} conditioned on $\mathbf{c}$, where $\mathbf{z}_t$ denotes the noisy placement state at diffusion step $t$, $\boldsymbol{\epsilon}_\theta(\mathbf{z}_t,\mathbf{c},t)$ is the predicted noise, and $\bar{\alpha}_t=\prod_{s=1}^{t}(1-\beta_s)$ is the cumulative signal retention coefficient defined by the forward variance schedule $\{\beta_s\}$. During reverse diffusion, the denoising network forms an estimate $\widehat{\mathbf{z}}_{0,\theta}$ of the clean macro placement $\mathbf{z}_0$ as
\begin{equation}
	\widehat{\mathbf{z}}_{0,\theta}
	=
	\frac{
		\mathbf{z}_t
		-
		\sqrt{1-\bar{\alpha}_t}\,
		\boldsymbol{\epsilon}_\theta
		(\mathbf{z}_t,\mathbf{c},t)
	}{
		\sqrt{\bar{\alpha}_t}
	}.
	\label{eq:predicted_placement_estimation}
\end{equation}
Prior diffusion guidance methods evaluate external objectives on a predicted denoised sample and map the resulting correction back to the noisy trajectory~\citep{bansal2024universal}. EGGPlacer applies this general principle to physical placement energies. Rather than evaluating them directly on the noisy state $\mathbf z_t$, it evaluates and optimizes them on the predicted placement $\widehat{\mathbf z}_{0,\theta}$, whose coordinates have explicit geometric meaning. This mechanism gives the macro placement generator access to a full netlist wirelength proxy and predicted temperature signals during inference.

To apply these signals to the predicted placement, let $\mathbf{u}\in\mathbb{R}^{N_M\times 2}$ denote a candidate clean macro placement in the same normalized coordinate space as $\mathbf{z}_0$. At each reverse step, $\mathbf{u}$ is initialized from $\widehat{\mathbf{z}}_{0,\theta}$ and refined using the composite placement energy
\begin{equation}
	\begin{aligned}
		E_{\mathrm{EG}}(\mathbf{u},\mathbf{c},t)
		={}&
		\lambda_{\mathrm{ana}}(t)
		E_{\mathrm{ana}}(\mathbf{u},\mathbf{c})
		+
		\lambda_{\mathrm{leg}}(t)
		E_{\mathrm{leg}}(\mathbf{u},\mathbf{c})
		+
		\lambda_{\mathrm{th}}(t)
		E_{\mathrm{th}}(\mathbf{u},\mathbf{c}),
	\end{aligned}
	\label{eq:combined_placement_energy}
\end{equation}
where $E_{\mathrm{ana}}$ exposes a full netlist weighted-average wirelength (WA-WL) proxy evaluated using the analytical placer's standard cell response, $E_{\mathrm{leg}}$ captures macro legality and spatial competition between macros and standard cells, and $E_{\mathrm{th}}$ captures thermal behavior. Together, these energies provide guidance signals for wirelength and thermal optimization while accounting for placement feasibility.

\paragraph{Sampling interpretation.}
Once the downstream objectives have been expressed as differentiable energies, their effect on sampling can be interpreted as tilting the conditional distribution of the predicted placement toward lower physical cost:
\begin{equation}
	q_t(\mathbf u\mid\mathbf z_t,\mathbf c)
	\propto
	p_\theta(\mathbf u\mid\mathbf z_t,\mathbf c)
	\exp\!\left[-\gamma_t E_{\mathrm{EG}}(\mathbf u,\mathbf c,t)\right],
	\label{eq:energy_tilted_placement_distribution}
\end{equation}
where $\gamma_t$ controls the guidance strength. Under this formulation, $p_\theta$ provides the learned placement prior, while the exponential energy term reweights its predictions toward lower physical cost using analytical and thermal feedback.

\paragraph{Energy refinement of the predicted placement.}
Because the conditional density in \Eqref{eq:energy_tilted_placement_distribution} is not explicitly available, EGGPlacer realizes this principle through local optimization of the predicted placement. It initializes a short refinement from the detached prediction:
\begin{equation}
	\mathbf{u}_t^{(0)}
	=
	\operatorname{sg}
	\left[
	\widehat{\mathbf{z}}_{0,\theta}
	\right],
	\qquad
	\mathbf{u}_t^{(K)}
	=
	\operatorname{Optimizer}_{K}
	\left(
	\mathbf{u}_t^{(0)};
	E_{\mathrm{EG}},
	\mathbf{c},
	t
	\right).
	\label{eq:energy_local_optimization}
\end{equation}
Here, $K$ denotes the number of local energy refinement steps, and $\operatorname{sg}[\cdot]$ denotes an operator that blocks gradient propagation. Detaching the initialization decouples the local physical optimization from the denoising network. The resulting placement correction is
\begin{equation}
	\mathbf{g}_t
	=
	\mathbf{u}_t^{(K)}
	-
	\mathbf{u}_t^{(0)}.
	\label{eq:energy_displacement}
\end{equation}
We use deterministic denoising diffusion implicit model (DDIM) sampling~\citep{DDIM} and incorporate this correction into the reverse transition as
\begin{equation}
	\mathbf{z}_{t-1}
	=
	\sqrt{\bar{\alpha}_{t-1}}\,
	\left(
		\widehat{\mathbf{z}}_{0,\theta}
		+
		\mathbf{g}_t
	\right)
	+
	\sqrt{1-\bar{\alpha}_{t-1}}\,
	\boldsymbol{\epsilon}_{\theta}(\mathbf{z}_t,\mathbf{c},t).
	\label{eq:energy_guided_ddim}
\end{equation}

The reverse process therefore alternates denoising and physical optimization while leaving the pretrained diffusion model unchanged.

\subsection{Analytical Placement Guidance}
mHPWL cannot reveal how standard cells respond to a candidate macro layout. We therefore use an analytical placer as an online physical response model during reverse sampling. At reverse step $t$, we fix the denoised macro coordinates $\mathbf{u}$ and compute the standard cell response $\mathbf{s}_t$ using $K_{\mathrm{sc}}$ Nesterov steps initialized at $\mathbf{s}^{\mathrm{init}}_t$:
\begin{equation}
	\mathbf{s}_t
	=
	\mathcal{A}_{K_{\mathrm{sc}}}
	\left(
	\mathbf{s}^{\mathrm{init}}_t;
	\operatorname{sg}[\mathbf{u}],\mathbf{c}
	\right),
	\label{eq:sc_response}
\end{equation}
where $\mathcal{A}$ applies $K_{\mathrm{sc}}$ DREAMPlace Nesterov steps to the full netlist wirelength proxy and density objective in \Eqref{eq:global_placement}, with macros held fixed~\citep{dreamplaceDAC,dreamplaceTCAD}.

Given this response, the analytical energy evaluates full netlist WA-WL, with $\mathbf{x}^{F}$ denoting fixed terminal coordinates and $\gamma>0$ controlling smoothing:
\begin{equation}
	\label{eq:ana_wl_energy}
	E_{\mathrm{ana}}(\mathbf u;\mathbf{s}_t,\mathbf c)
	=
	\operatorname{WA\text{-}WL}_{\gamma}
	\left(
	\mathbf u,
	\operatorname{sg}[\mathbf{s}_t],
	\mathbf{x}^{F};E
	\right).
\end{equation}
Unlike mHPWL, \Eqref{eq:ana_wl_energy} includes pins and nets incident to standard cells, macros, and fixed terminals. When evaluating $E_{\mathrm{ana}}$, the standard cell coordinates $\mathbf{s}_t$ are treated as constants, while the macro coordinates remain variable. This avoids backpropagating through the analytical solver while closing the feedback loop by recomputing the response at the next reverse step.

The standard cell response also indicates space that should remain available to small cells. We represent its regions of high occupancy by a set of obstacle rectangles $\mathcal{R}_t$. Let $B_i(\mathbf u_i)$ denote the rectangle occupied by macro $i$, and let $B_j$ denote either the rectangle $B_j(\mathbf u_j)$ occupied by another macro or an obstacle rectangle in $\mathcal R_t$. We define
\begin{equation}
	E_{\mathrm{leg}}(\mathbf u;\mathcal R_t)
	=
	\sum_{(i,j)\in\mathcal P_t}
	\widetilde{A}_{\cap}
	\left(B_i(\mathbf u_i),B_j\right)
	+
	\lambda_{\mathrm{bd}}E_{\mathrm{boundary}}(\mathbf u),
	\label{eq:aux_feasibility_energy}
\end{equation}
where $\widetilde{A}_{\cap}(B_i,B_j)$ denotes the smooth overlap between $B_i$ and $B_j$, $\mathcal P_t$ contains unordered pairs from the union of macros and obstacles, $E_{\mathrm{boundary}}$ denotes the smooth penalty for macro boundary violations, and $\lambda_{\mathrm{bd}}$ is its weight. When evaluating this energy term, obstacle coordinates are treated as constants, whereas macro coordinates remain variable. At each reverse step, EGGPlacer first obtains the response in \Eqref{eq:sc_response} and then jointly refines macros using the single composite energy in \Eqref{eq:combined_placement_energy}.

\subsection{Thermal Guidance}
Accurate thermal simulation is too expensive to invoke inside every reverse step. We therefore freeze a differentiable surrogate $F_{\psi}$ that operates on sets during sampling. It maps the macro coordinates $\mathbf u$, together with static geometry, each macro's type (i.e., its library master), and power context, to temperature estimates $\widehat{\mathbf t}$ for individual macros. The surrogate also includes a global head that predicts chip-level temperature statistics for calibration analysis, while the per-macro predictions define the guidance objective.

To capture both the overall macro temperature and localized hotspots, we combine a mean temperature weighted by macro area with a smooth maximum. Let $a_i$ denote the normalized area of macro $i$ and $\widehat t_i$ its temperature predicted by the thermal surrogate $F_{\psi}$. We define
\begin{equation}
	T_{\mathrm{area}}
	=
	\frac{\sum_{i=1}^{N_M}a_i\widehat t_i}
	{\sum_{i=1}^{N_M}a_i},
	\qquad
	T_{\mathrm{hot}}
	=
	\tau_T\log\!\left(
	\frac{1}{N_M}
	\sum_{i=1}^{N_M}
	\exp\!\left(\frac{\widehat t_i}{\tau_T}\right)
	\right),
	\label{eq:thermal_macro_statistics}
\end{equation}
where $T_{\mathrm{area}}$ is the mean of the predicted macro temperatures weighted by area, $T_{\mathrm{hot}}$ is their smooth maximum, and $\tau_T$ controls the smoothness of this maximum. The thermal guidance energy is
\begin{equation}
	E_{\mathrm{th}}(\mathbf u,\mathbf c_{\mathrm{th}})
	=
	(1-\rho)T_{\mathrm{area}}+\rho T_{\mathrm{hot}},
	\label{eq:thermal_base_energy}
\end{equation}
where $\rho$ balances the mean weighted by macro area and the smooth maximum. Because the surrogate and the thermal objective are differentiable with respect to the macro coordinates, automatic differentiation provides the coordinate information required for thermal guidance.

The thermal term participates in the same optimization of the predicted placement as the wirelength proxy and legality. Its coefficient is scaled relative to the nonthermal terms and reduced when the current placement is infeasible or when the thermal objective conflicts with legality. This makes thermal guidance a bounded local preference rather than an unconstrained postprocessing displacement. The surrogate is used only to provide differentiable thermal guidance during sampling.

\paragraph{Macro legalization.}
After the final reverse step, we apply a deterministic greedy legalizer on a
$256\times256$ grid to remove residual macro overlaps and boundary violations.
Macros that are already legal remain fixed. The remaining macros are processed
in descending order of area and moved to the closest feasible grid point. All
reported EGGPlacer metrics are computed after applying this procedure, which
leaves placements without residual violations unchanged.

\section{Experiments}
\label{Experiments}

\subsection{Experimental Setup}
\label{sec:experimental_setup}

We evaluate EGGPlacer on 20 designs from the open-source ChiPBench benchmark suite~\citep{chipbench}, which supports an end-to-end physical design flow with OpenROAD and is based on the Nangate $45\,\mathrm{nm}$ technology library~\citep{FreePDK}.
Our denoiser combines graph and global attention with timestep conditioning and is trained from scratch on the ChipDiffusion \texttt{v1} synthetic dataset~\citep{chipdiffusion}.
DREAMPlace~4.2~\citep{dreamplaceDAC,dreamplaceTCAD} provides the analytical placement engine for EGGPlacer guidance. Downstream physical design and thermal evaluation use the OpenROAD Flow Scripts (ORFS)~\citep{kahng2021openroad} and HotSpot~\citep{skadron2004temperature,han2021hotspot7}, respectively.
For HotSpot, both the ambient and initial temperatures are set to $318.15\,\mathrm{K}$ ($45\,^{\circ}\mathrm{C}$).
All experiments are conducted on a single workstation equipped with an NVIDIA GeForce RTX~2080~Ti GPU and an Intel Xeon Gold~6226R CPU running at 2.90~GHz.
For EGGPlacer and each ablation variant, we report results averaged over three random seeds.

We compare EGGPlacer against five representative placement approaches: EfficientPlace (EffPlace)~\citep{efficientplace}, ChipDiffusion (ChipDiff)~\citep{chipdiffusion}, RollPlace~\citep{RollPlace}, FlowPlace~\citep{flowplace}, and DREAMPlace (DP)~\citep{dreamplaceDAC,dreamplaceTCAD}.
LaMPlace~\citep{lamplace_crossstage} is not included because we could not reliably reproduce its reported results with the currently available implementation under our evaluation pipeline.
DREAMPlace is additionally evaluated in its native mixed-size placement mode as a strong analytical baseline.
All baselines and downstream electronic design automation (EDA) components use open-source implementations.

We evaluate placement quality at both placement and routing stages. Wirelength estimates, referred to here as wirelength proxies, provide inexpensive indicators of routed interconnect quality~\citep{caldwell1998wirelength,kahng2003accurate}. Among the reported HPWL metrics, mHPWL includes only macros and fixed terminals, whereas mixed-size HPWL also includes standard cells. We denote mixed-size HPWL after the common DREAMPlace refinement by $\text{HPWL}_{\text{DP}}$ and HPWL after the ORFS placement stage by $\text{HPWL}_{\text{ORFS}}$. Routed wirelength denotes the total interconnect length reported after ORFS routing.
Thermal quality is characterized using the peak temperature estimated by HotSpot.
Unless otherwise specified, lower values indicate better performance.

\subsection{Placement Quality and Downstream Evaluation}

\paragraph{Mixed-size placement quality.}
Unlike previous learning-based macro placers that primarily optimize macro-level objectives, EGGPlacer injects analytical placement information into the diffusion process, allowing macro decisions to account for the spatial requirements of the surrounding standard cells.
To quantify this effect, we construct a controlled two-stage evaluation protocol.
First, each macro placement method determines the macro locations.
Second, these macro locations are fixed, and DREAMPlace is used with an identical configuration to optimize the remaining standard cells.
This protocol allows us to evaluate whether a macro placement provides a favorable initialization for subsequent mixed-size placement, rather than measuring the mHPWL proxy alone.
All macro placements are legal except the ChipDiffusion result for or1200, which contains a macro overlap and is therefore excluded from its aggregate statistics.

\autoref{tab:DP_benchmark_comparison} reports mHPWL and the DREAMPlace-refined mixed-size proxy $\text{HPWL}_{\text{DP}}$ on all 20 benchmarks. Among the learning-based macro placement methods, EGGPlacer achieves a competitive average $\text{HPWL}_{\text{DP}}$, closely matching RollPlace while outperforming FlowPlace.
DREAMPlace still obtains the lowest $\text{HPWL}_{\text{DP}}$, which is expected because its objective directly targets a differentiable mixed-size wirelength proxy and density throughout optimization.
Importantly, however, the rankings under mHPWL and $\text{HPWL}_{\text{DP}}$ differ considerably across the evaluated methods.
These differing rankings show that mHPWL does not fully reflect mixed-size placement quality. Macro configurations that appear favorable when standard cells are ignored can leave insufficient or poorly distributed whitespace for subsequent standard cell placement.
This limitation motivates incorporating mixed-size physical information directly into the macro generation process, as done by EGGPlacer.

\begin{table}[htbp]
	\centering
	\caption{Mixed-size placement summary across 20 benchmarks. The macro-only
		proxy mHPWL and the DREAMPlace-refined proxy $\text{HPWL}_{\text{DP}}$
		are reported in units of $10^{8}$ and $10^{9}$, respectively. DP,
		EffPlace, and ChipDiff denote DREAMPlace, EfficientPlace, and
		ChipDiffusion, respectively. EGGPlacer
		values are arithmetic means over three random seeds. Lower is better.}
	\label{tab:DP_benchmark_comparison}
	\begin{tabular}{l cc cc}
		\toprule
		\multirow{2}{*}{Method}
		& \multicolumn{2}{c}{mHPWL}
		& \multicolumn{2}{c}{$\text{HPWL}_{\text{DP}}$} \\
		\cmidrule(lr){2-3}\cmidrule(lr){4-5}
		& Mean & Rank & Mean & Rank \\
		\midrule
		DP & 1.46 & 3.75 & \textbf{5.64} & \textbf{1.65} \\
		EffPlace + DP & \underline{1.00} & \underline{1.55} & 6.69 & 4.20 \\
		ChipDiff + DP & \textbf{0.90} & \textbf{1.53} & 6.89 & 4.16 \\
		RollPlace + DP & 1.74 & 5.10 & \underline{6.05} & \underline{2.95} \\
		FlowPlace + DP & 1.68 & 3.85 & 6.82 & 4.65 \\
		EGGPlacer + DP & 1.71 & 5.00 & 6.08 & 3.30 \\
		\bottomrule
	\end{tabular}
	\vspace{-3mm}
\end{table}

\paragraph{Runtime.}
\autoref{tab:average_runtime_comparison} summarizes the average runtime over the 20 benchmarks, decomposed into macro placement, mixed-size placement, and total runtime.
EGGPlacer combines competitive placement quality with moderate runtime among the learning-based macro placers.
EGGPlacer completes macro placement and mixed-size placement in $95.40$ seconds, remaining substantially faster than EfficientPlace, ChipDiffusion, and RollPlace. FlowPlace is the fastest learning-based baseline, although this speed does not translate into a correspondingly strong $\text{HPWL}_{\text{DP}}$. The high runtime of RollPlace reflects its search-based optimization procedure.


\begin{table}[htbp]
	\centering
	\caption{Average runtime across 20 benchmarks. DP,
		EffPlace, and ChipDiff denote DREAMPlace, EfficientPlace, and
		ChipDiffusion, respectively. Mixed-size placement refers to the DP refinement stage. EGGPlacer runtimes are averaged over three
		random seeds.}
	\label{tab:average_runtime_comparison}
	
	\begin{tabular}{l c c c c c c}
		\toprule
		Runtime (s)
		& DP
		& EffPlace
		& ChipDiff
		& RollPlace
		& FlowPlace
		& EGGPlacer \\
		\midrule
		
		Macro Placement
		& -- & 769.66 & 128.29 & 8802.57 & 6.40 & 79.14 \\
		
		Mixed-Size Placement
		& 19.73 & 18.66 & 18.51 & 16.06 & 17.88 & 16.26 \\
		
		\midrule
		
		Total
		& 19.73
		& 788.32
		& 146.80
		& 8818.63
		& 24.28
		& 95.40 \\
		
		\bottomrule
	\end{tabular}
	\vspace{-2mm}
\end{table}

\paragraph{End-to-end physical design evaluation.}
The $\text{HPWL}_{\text{DP}}$ proxy alone does not fully characterize the quality of a macro placement for downstream physical design.
We therefore evaluate the macro placements generated in the preceding mixed-size experiment using a complete EDA flow.
With macro locations fixed, ORFS completes the subsequent placement and routing stages, after which HotSpot estimates the thermal distribution of each completed design.
This evaluation directly examines whether the physical guidance introduced during generation produces macro configurations that remain advantageous after downstream optimization.

\autoref{tab:openroad_benchmark_comparison} reports ORFS HPWL, routed wirelength, and peak temperature obtained from the complete OpenROAD and HotSpot evaluation.
EGGPlacer achieves the lowest averages across all three metrics. For the two downstream metrics central to this evaluation, it reduces routed wirelength by 2.18\% relative to RollPlace and peak temperature rise by 4.90\% relative to ChipDiffusion, the strongest external baseline for each metric.
Notably, although DREAMPlace achieves lower $\text{HPWL}_{\text{DP}}$ in the preceding mixed-size placement experiment, this advantage does not consistently translate into superior routed results after the complete physical design flow.
This discrepancy highlights the limited fidelity of HPWL as a predictor of routed wirelength.

\begin{table}[htbp]
	\centering
	\caption{End-to-end summary across 20 benchmarks. $\text{HPWL}_{\text{ORFS}}$
		and routed wirelength are reported in $10^{6}\,\mu\mathrm{m}$ and peak temperature in
		$^{\circ}\mathrm{C}$. DP, EffPlace, and ChipDiff denote DREAMPlace,
		EfficientPlace, and ChipDiffusion, respectively. EGGPlacer values are arithmetic means over three
		random seeds, while external baselines report single runs. Lower is better.}
	\label{tab:openroad_benchmark_comparison}
	\begin{tabular}{l cc cc cc}
		\toprule
		\multirow{2}{*}{Method}
		& \multicolumn{2}{c}{$\text{HPWL}_{\text{ORFS}}$}
		& \multicolumn{2}{c}{Routed WL}
		& \multicolumn{2}{c}{Peak T} \\
		\cmidrule(lr){2-3}\cmidrule(lr){4-5}\cmidrule(lr){6-7}
		& Mean & Rank & Mean & Rank & Mean & Rank \\
		\midrule
		DP & 4.31 & 3.40 & 5.48 & 3.25 & 62.93 & 3.55 \\
		EffPlace & 4.88 & 4.40 & 5.86 & 4.40 & 65.38 & 5.30 \\
		ChipDiff & 4.86 & 4.53 & 5.85 & 4.53 & \underline{62.35} & \underline{3.11} \\
		RollPlace & \underline{4.21} & 3.05 & \underline{5.10} & 3.10 & 63.18 & 3.40 \\
		FlowPlace & 4.40 & \underline{2.95} & 5.26 & \underline{2.85} & 63.63 & 3.70 \\
		EGGPlacer & \textbf{4.13} & \textbf{2.60} & \textbf{4.99} & \textbf{2.80} & \textbf{61.50} & \textbf{1.75} \\
		\bottomrule
	\end{tabular}
\end{table}

The thermal results reveal an additional limitation of optimization driven purely by a wirelength proxy.
Because thermal behavior is absent from the DREAMPlace objective, aggressive proxy minimization can favor spatially compact macro configurations that are undesirable from a thermal perspective.
In contrast, EGGPlacer incorporates thermal guidance directly into macro generation. Despite a higher average $\text{HPWL}_{\text{DP}}$ than DREAMPlace, it achieves lower routed wirelength and peak temperature after the end-to-end flow, demonstrating the value of exposing complementary downstream objectives beyond a single wirelength proxy.
To keep the primary comparison tables focused on the complete method and external baselines, we evaluate the variants with only analytical or thermal guidance separately in the following ablation study.

\subsection{Analysis and Ablation}

\begin{figure}[htbp]
	\centering
	\begin{subfigure}[t]{0.49\linewidth}
		\centering
		\includegraphics[width=\linewidth]{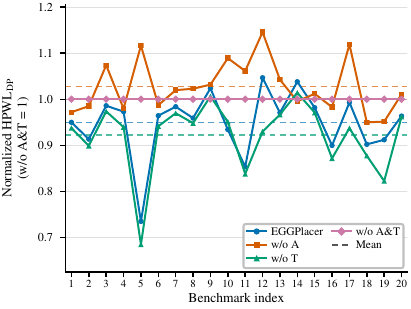}
		\caption{$\text{HPWL}_{\text{DP}}$ proxy after DREAMPlace refinement.}
		\label{fig:ablation-hpwl}
	\end{subfigure}
	\hfill
	\begin{subfigure}[t]{0.49\linewidth}
		\centering
		\includegraphics[width=\linewidth]{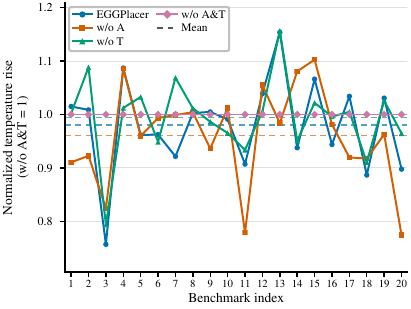}
		\caption{HotSpot peak temperature.}
		\label{fig:ablation-peak-temperature}
	\end{subfigure}

	\caption{Ablation of analytical and thermal guidance. Means over three seeds
		are normalized per benchmark to EGGPlacer w/o A\&T. Lower is better.
		Benchmark indices from 1 to 20 correspond, respectively, to ariane133, ariane136,
		ariane81, bp, bp68, bp\_be, bp\_be12, bp\_fe, bp\_fe38, bp\_multi,
		bp\_multi57, dft68, ethernet, isa\_npu, mor1kx, or1200, swerv\_wrapper,
		swerv\_wrapper43, vga\_lcd, and VeriGPU.
		Peak temperature is normalized as the rise above $318.15\,\mathrm{K}$ ($45\,^{\circ}\mathrm{C}$).
		Markers denote benchmarks, solid lines connect each variant's results,
		and dashed lines denote variant means.}
	\label{fig:energy-guidance-ablation}
	\vspace{-6mm}
\end{figure}

\paragraph{Effect of energy guidance.}
We isolate the analytical (A) and thermal (T) energy terms by comparing four variants: w/o A\&T removes both guidance terms, w/o A retains only T, w/o T retains only A, and the complete EGGPlacer uses both.
Figure~\ref{fig:energy-guidance-ablation} compares the variants using the $\text{HPWL}_{\text{DP}}$ proxy obtained after DREAMPlace refinement and HotSpot peak temperature, representing placement and thermal quality, respectively.
We use the same guidance hyperparameters across all designs, with energy weights adapted during sampling.

Adding analytical guidance produces the strongest average reduction in $\text{HPWL}_{\text{DP}}$, showing that the analytical response favors macro configurations that better accommodate subsequent standard cell placement. Adding thermal guidance instead produces the strongest average reduction in peak temperature rise, confirming that it steers generation away from thermally unfavorable configurations. Relative to the unguided variant, the complete EGGPlacer reduces the mean normalized $\text{HPWL}_{\text{DP}}$ by 5.1\% and peak temperature rise by 2.0\%. The complete configuration therefore balances wirelength reduction and thermal improvement through joint analytical and thermal guidance.

\paragraph{Engine dependence of placement proxies for routed wirelength.}
DREAMPlace and the RePlAce-based global placer in ORFS share an electrostatic analytical formulation with smooth wirelength and density objectives~\citep{nonlinear7eplace,nonlinear8replace,dreamplaceDAC,kahng2021openroad}. These shared foundations might suggest that the HPWL values produced by the two placement flows are interchangeable. However, the two proxies are measured after different standard cell placement flows with the macro locations fixed. We therefore examine whether $\text{HPWL}_{\text{DP}}$ and $\text{HPWL}_{\text{ORFS}}$ rank the same macro placements consistently and how closely each agrees with routed wirelength.

The controlled comparison uses the four guidance variants on all 20 benchmarks. For each design and variant, metrics are first averaged over three seeds and then normalized within the design to remove differences in circuit scale. Spearman correlation measures the monotonic association of each proxy with routed wirelength, while pairwise rank agreement measures how often the proxy and routed wirelength order two variants consistently within the same design. Table~\ref{tab:engine-dependent-proxy-summary} summarizes the results.

\begin{table}[htbp]
	\centering
	\caption{Fidelity of HPWL across placement engines for 80
		observations covering all combinations of variants and designs. Each observation is
		averaged over three random seeds and normalized within its design. Rank
		agreement with routed wirelength compares pairwise orderings.}
	\label{tab:engine-dependent-proxy-summary}
	\begin{tabular}{lcc}
		\toprule
		Placement metric & Spearman $\rho$ (95\% CI) & Rank agreement \\
		\midrule
		$\text{HPWL}_{\text{DP}}$ & $0.660$ $[0.405,0.842]$ & $72.5\%$ \\
		$\text{HPWL}_{\text{ORFS}}$ & $0.918$ $[0.817,0.977]$ & $91.7\%$ \\
		\bottomrule
	\end{tabular}
\end{table}

Table~\ref{tab:engine-dependent-proxy-summary} shows that $\text{HPWL}_{\text{ORFS}}$ has a substantially stronger Spearman correlation with routed wirelength than $\text{HPWL}_{\text{DP}}$, increasing from $0.660$ to $0.918$. Pairwise rank agreement likewise increases from $72.5\%$ to $91.7\%$, showing that the stronger association also holds when comparing placement configurations within individual designs. Although the two proxies share similar analytical foundations, their fidelity to routed wirelength can vary across placement engines. The stronger ORFS association is consistent with its alignment to the target physical design flow and helps explain why lower $\text{HPWL}_{\text{DP}}$ does not consistently translate into shorter routed wirelength.

\FloatBarrier
\section{Conclusion}
\label{Conclusion}
We presented EGGPlacer, an energy-guided diffusion framework that exposes downstream physical objectives during macro generation. Analytical guidance provides standard cell response and a full netlist WA-WL proxy, while thermal guidance supplies a differentiable hotspot objective. On 20 open-source designs, EGGPlacer achieves strong combined results in HPWL, routed wirelength, thermal performance, and runtime without retraining its diffusion backbone. We also show that the fidelity of HPWL as a proxy for routed wirelength can vary across placement engines, motivating proxies aligned with the target downstream flow. Promising directions include improving the calibration of physical predictors and developing differentiable proxies for additional downstream quantities, such as worst negative slack (WNS) and total negative slack (TNS).

\subsection*{AI use statement}

In this work, we used generative AI tools to refine the research methodology and experimental design, formulate mathematical descriptions, present experimental results, edit the manuscript for clarity and organization, and assist with code development. We independently reviewed and validated all mathematical formulations and experimental protocols developed with generative AI assistance and reviewed and tested all AI-assisted code. We take responsibility for the final content of this work, including all text, claims, results, and artifacts produced with the aid of generative AI.

\bibliography{iclr2027_conference}
\bibliographystyle{iclr2027_conference}

\appendix
\section{Related Work}
\label{Sec:RelatedWork}

\subsection{Optimization-Based Placement}
\label{subSec:OptimizationBasedMethods}

Optimization-based macro placement is commonly approached through simulated
annealing and analytical optimization.

\textbf{Methods based on simulated annealing.}
Macro placement can be formulated as a combinatorial black-box optimization problem.
Simulated annealing (SA) explores this space through iterative perturbation and evaluation.
Recent methods further exploit hierarchy and dataflow information in RTL designs.
RTL-MP~\citep{RTL-MP} clusters RTL netlists while preserving their structural relationships and optimizes the resulting clusters using simulated annealing.
Hier-RTLMP~\citep{Hier-RTLMP2024} extends this idea through hierarchical and multiscale optimization and has been integrated into the OpenROAD ecosystem~\citep{kahng2021openroad} for reproducible evaluation from RTL to GDSII.

\textbf{Analytical methods.}
Analytical placers relax discrete placement into continuous optimization by expressing wirelength proxies and density constraints as differentiable functions of cell coordinates.
Modern nonlinear placers use smooth wirelength approximations and density penalties to achieve high-quality solutions~\citep{nonlinear6,nonlinear7eplace,nonlinear8replace}.
DREAMPlace~\citep{dreamplaceDAC,dreamplaceTCAD}, building on NTUplace3~\citep{nonlinear6}, ePlace~\citep{nonlinear7eplace}, and RePlAce~\citep{nonlinear8replace}, reformulates analytical placement using tensor operations and accelerates optimization on GPUs with PyTorch.
AutoDMP~\citep{AutoDMP} combines DREAMPlace with Bayesian optimization to tune macro placement parameters, while Xplace~\citep{Xplace} further improves computational efficiency through optimized GPU operators.

\subsection{Learning-Based Macro Placement}
\label{subSec:LearningBasedMethods}

Learning-based methods have recently emerged as an alternative paradigm for macro placement, aiming to learn placement policies or structural priors directly from circuit data.
Google's pioneering work~\citep{mirhoseini2021graph} formulates macro placement as a Markov Decision Process and sequentially places macros using reinforcement learning (RL).
MaskPlace~\citep{maskplace} improves learning efficiency through view, position, and wire masks, where the wire mask provides dense feedback for evaluating candidate macro locations.
PRNet~\citep{cheng2022policy} further incorporates macro overlap into the optimization objective to encourage feasible placements, while WireMask-BBO~\citep{wiremaskbbo} transfers the wire mask representation of MaskPlace to black-box optimization for more effective exploration.
Recent work has also extended the placement action space itself. For example, RSPlace~\citep{rsplace} explicitly considers macro orientation and jointly searches over placement and rotation decisions through bidirectional tree expansion.

A major challenge for RL-based macro placement is the substantial cost of training, adaptation, and sequential exploration.
ChiPFormer~\citep{chipformer} addresses cross-circuit generalization using offline reinforcement learning with a Decision Transformer~\citep{Ashish2017Attention}, enabling knowledge transfer with lightweight online adaptation.
EfficientPlace~\citep{efficientplace} combines learned local policies with global tree search to improve sample efficiency, while Monte Carlo tree search (MCTS) guided by a pretrained RL policy~\citep{mctsrlplace} further exploits learned policies to accelerate macro placement search on large designs.
Rather than using RL exclusively as a standalone placer, MacroRegulator~\citep{macroregulator} employs a learned policy to regulate and improve existing macro placements, highlighting the effectiveness of learning-based refinement.
LAMPlace~\citep{lamplace_legalization} incorporates legalization into RL-based placement for mixed-size designs with preplaced blocks.
Similarly, RollPlace~\citep{RollPlace} refines initial macro configurations through Monte Carlo rollout search, decoupling the quality of the final placement from a single sequential placement policy.

Beyond improving search efficiency, recent studies have increasingly recognized the mismatch between macro-level proxy objectives and downstream physical design quality.
LaMPlace~\citep{lamplace_crossstage} learns surrogate models of placement metrics and converts their predictions into masks that guide macros to be placed sequentially, one at a time.
Other approaches incorporate domain expertise into the learning process.
Expertise-based RL~\citep{expertise_macro} introduces physical design knowledge and expert preferences into macro placement policies, while subsequent work on expert-level placement~\citep{expertlevel_rl} leverages expert placement information to guide RL toward higher-quality solutions.
These methods demonstrate that placement quality can benefit substantially from information that is not captured by conventional macro-level wirelength objectives alone.

Generative models provide another direction for learning-based placement.
ChipDiffusion~\citep{chipdiffusion} formulates continuous coordinate macro placement as a conditional diffusion process, replacing autoregressive macro-by-macro decisions with joint generative sampling.
More recently, FlowPlace~\citep{flowplace} employs flow matching~\citep{lipman2023flowmatchinggenerativemodeling} to construct a continuous generative trajectory for chip placement, providing an alternative to diffusion-based generation.
Such generative formulations offer increased flexibility for incorporating additional constraints and objectives during inference.

Despite these advances, a fundamental challenge remains in aligning the objective used by an AI-based macro placer with the quality of the resulting downstream physical design.
Earlier RL methods predominantly optimize inexpensive macro-level proxies such as mHPWL, whereas more recent approaches mitigate this mismatch through learned cross-stage predictors, expert knowledge, or demonstrations.
However, these signals are typically learned offline or encoded indirectly and do not directly couple the generative trajectory with online feedback from a downstream analytical placement process.
Moreover, thermal characteristics, which depend on both placement geometry and power distribution, are rarely exposed to the macro generation process.

EGGPlacer addresses this gap by adopting a diffusion-based generative backbone while introducing analytical placement and thermal objectives as energy guidance during inference.
Instead of relying solely on macro-level surrogate objectives or separately learned placement policies, EGGPlacer evaluates physical objectives on intermediate denoised placements and uses their gradients to steer the reverse diffusion trajectory.
This formulation exposes downstream standard cell placement behavior and thermal information directly during generation, enabling multiple physical objectives to jointly influence the resulting macro placement.

\section{End-to-End Physical Design Evaluation}
\label{sec:end-to-end-evaluation-details}

Following ChiPBench, we keep the generated macros fixed and evaluate all methods
with the same ORFS configuration. \textbf{Global Placement} optimizes standard
cells around the macros. \textbf{Legalization} moves cells to legal row sites
and removes overlaps. \textbf{Clock Tree Synthesis (CTS)} constructs the clock
distribution network. \textbf{Routing} combines global routing, which generates
interconnect guides, with detailed routing, which realizes legal interconnects.
Further details are provided by ChiPBench and OpenROAD~\citep{chipbench,kahng2021openroad}.

\section{Diffusion Backbone and Training Details}
\label{sec:diffusion-backbone-details}

\paragraph{Backbone and representation.}
EGGPlacer follows the continuous coordinate diffusion formulation and synthetic data protocol of ChipDiffusion~\citep{chipdiffusion}.
The denoiser conditions on normalized noisy macro centers, macro sizes, netlist features that encode pin information, fixed ports, and the timestep.
An encoder with width 64 feeds two hybrid blocks that combine GATv2 message passing~\citep{gat} with global attention using four heads~\citep{Ashish2017Attention} and timestep conditioning. The decoder predicts Gaussian noise in two dimensions for each movable node.
The model has 556,354 trainable parameters. All EGGPlacer variants use the same frozen checkpoint, so their comparison isolates the guidance applied during inference.

\paragraph{Data and training.}
We train from scratch on the ChipDiffusion~\citep{chipdiffusion} \texttt{v1} synthetic dataset, which contains 40,000 training and 2,000 validation circuits. No samples from real designs or the \texttt{v2} fine-tuning set are used.
Training uses a cosine diffusion schedule with 200 steps and the masked noise prediction MSE in \Eqref{eq:diffusion_training}. We run Adam for 3,000,000 updates with batch size 64, peak learning rate $10^{-4}$, and EMA decay 0.999. Deterministic DDIM sampling uses $\eta=0$.

\subsection{Diffusion Objective and Predicted Placement Guidance Derivation}
\label{sec:predicted-placement-guidance-derivation}

For completeness, let
$\alpha_t=1-\beta_t$ and
$\bar\alpha_t=\prod_{s=1}^{t}\alpha_s$.  The standard forward process is
\begin{equation}
\mathbf z_t
=\sqrt{\bar\alpha_t}\,\mathbf z_0
+\sqrt{1-\bar\alpha_t}\,\boldsymbol\epsilon,
\qquad
\boldsymbol\epsilon\sim\mathcal N(\mathbf 0,\mathbf I),
\label{eq:forward_diffusion}
\end{equation}
and the denoiser is trained using
\begin{equation}
\mathcal L_{\mathrm{diff}}(\theta)
=\mathbb E_{\mathbf z_0,t,\boldsymbol\epsilon}
\!\left[
\left\|\boldsymbol\epsilon-
\boldsymbol\epsilon_\theta(\mathbf z_t,\mathbf c,t)\right\|_2^2
\right].
\label{eq:diffusion_training}
\end{equation}

Starting from the distribution tilted by the energy in
\Eqref{eq:energy_tilted_placement_distribution}, we approximate the unavailable
conditional negative log density locally around the predicted placement by a
quadratic trust region:
\begin{equation}
\mathbf u_t^*
=\arg\min_{\mathbf u}
\left[
\frac{1}{2\tau_t}
\left\|\mathbf u-\operatorname{sg}
[\widehat{\mathbf z}_{0,\theta}]\right\|_2^2
+\gamma_t E_{\mathrm{EG}}(\mathbf u,\mathbf c,t)
\right].
\label{eq:local_energy_inference}
\end{equation}
The first term preserves the local diffusion prior, whereas the second favors
physical objectives.  EGGPlacer realizes this local inference with the finite
optimization in \Eqref{eq:energy_local_optimization}.  With one gradient
step, its displacement reduces to the familiar gradient update
\begin{equation}
\mathbf g_t
\approx
-\eta_t\nabla_{\mathbf u}E_{\mathrm{EG}}(\mathbf u,\mathbf c,t)
\big|_{\mathbf u=\mathbf u_t^{(0)}}.
\label{eq:single_step_energy_guidance}
\end{equation}
Multiple inner steps permit structured physical energies to refine the same
predicted placement before it is returned to the diffusion trajectory.

For deterministic DDIM sampling~\citep{DDIM}, substituting the refined
prediction $\widehat{\mathbf z}_{0,\theta}+\mathbf g_t$ into the reverse step
gives \Eqref{eq:energy_guided_ddim}. Setting $\mathbf g_t=0$ recovers the
standard unguided DDIM transition.

\section{Inference and Guidance Implementation Details}
\label{sec:guidance-implementation-details}

This section details the numerical mechanisms used for guided inference.

\subsection{Analytical Placement Guidance}
\label{sec:analytical-guidance-details}

At each reverse step, the analytical placer performs $K_{\mathrm{sc}}=5$
Nesterov steps to update the standard cells while keeping the macros fixed.
The resulting standard cell coordinates are treated as constants when
evaluating the full netlist WA-WL energy. For the evaluated net
set $\mathcal{N}$, let $\mathcal{P}_e$ be the pins of net $e$ and $q_{i,d}$ the absolute coordinate of pin $i$
along dimension $d\in\{x,y\}$, including its pin offset. We define
\begin{equation}
\begin{aligned}
\operatorname{WA}_{\gamma}(e,d)
&=
\frac{\sum_{i\in\mathcal{P}_e}q_{i,d}\exp(q_{i,d}/\gamma)}
{\sum_{i\in\mathcal{P}_e}\exp(q_{i,d}/\gamma)}
-
\frac{\sum_{i\in\mathcal{P}_e}q_{i,d}\exp(-q_{i,d}/\gamma)}
{\sum_{i\in\mathcal{P}_e}\exp(-q_{i,d}/\gamma)},\\
\operatorname{WA\text{-}WL}_{\gamma}
&=\sum_{e\in\mathcal{N}}w_e
\sum_{d\in\{x,y\}}\operatorname{WA}_{\gamma}(e,d).
\end{aligned}
\label{eq:weighted-average-wirelength}
\end{equation}
Here $w_e$ is the net weight, and $\gamma>0$ controls the smoothing. Smaller
values of $\gamma$ yield a closer approximation to HPWL. Consequently,
the macro gradient used by guidance is
\begin{equation}
\nabla_{\mathbf u}E_{\mathrm{ana}}
=
\left.
\frac{\partial\operatorname{WA\text{-}WL}_{\gamma}}
{\partial\mathbf u}
\right|_{\mathbf{s}=\operatorname{sg}[\mathbf{s}_t]}.
\label{eq:detached_analytical_gradient}
\end{equation}
To obtain the standard cell response in \Eqref{eq:sc_response}, we refine the
standard cell coordinates with the DREAMPlace Nesterov optimizer while holding
the macros fixed, using its GPU implementations of the WA-WL and electric
potential density operators. The analytical energy for macro guidance then evaluates and
differentiates only the full netlist WA-WL term with the same DREAMPlace
wirelength operator, treating the refined standard cell coordinates as constants.

To construct the auxiliary feasibility term, we form a $64\times64$ occupancy
histogram weighted by standard cell area and mark bins whose occupancy exceeds
$0.1$. We identify connected components using connectivity among the four
orthogonal neighbors and greedily decompose each component into rectangles
aligned with the coordinate axes. Rectangles smaller than $0.1\%$ of the chip
area are discarded. The remaining rectangles are concatenated with the macro
boxes before pairwise overlap is evaluated. For boxes $B_i$ and $B_j$, let
$(l_i,r_i,b_i,q_i)$ and $(l_j,r_j,b_j,q_j)$ denote their left, right, bottom,
and top boundaries. Their smooth overlap is
\begin{equation}
\widetilde{A}_{\cap}(B_i,B_j)
=
\operatorname{sp}_{\beta}
\!\left(\min(r_i,r_j)-\max(l_i,l_j)\right)
\operatorname{sp}_{\beta}
\!\left(\min(q_i,q_j)-\max(b_i,b_j)\right),
\label{eq:smooth_rectangle_overlap}
\end{equation}
where
$\operatorname{sp}_{\beta}(x)=\beta^{-1}\log(1+\exp(\beta x))$, with
$\beta$ set to 100. When evaluating the smooth overlap energy, the standard
cell rectangles are treated as constants, while the macro boxes remain
variable.

The gradients of the legality and WA-WL energies can differ substantially in scale. Define
\begin{equation}
G_{\mathrm{ana}}=\operatorname{RMS}(\nabla_{\mathbf u}E_{\mathrm{ana}}),
\qquad
G_{\mathrm{ref}}
=\max(\lambda_{\mathrm{leg}},\lambda_{\mathrm{floor}})
\operatorname{RMS}(\nabla_{\mathbf u}E_{\mathrm{leg}}).
\end{equation}
The detached target WA-WL coefficient is
\begin{equation}
\lambda_{\mathrm{ana}}^{\mathrm{tar}}
=
\operatorname{clip}\!\left(
r_{\mathrm{ana}}\frac{G_{\mathrm{ref}}}{G_{\mathrm{ana}}+\epsilon},
\lambda_{\mathrm{ana}}^{\min},
\lambda_{\mathrm{ana}}^{\max}
\right).
\label{eq:analytical_rms_balance}
\end{equation}
We use $r_{\mathrm{ana}}=0.4$, an EMA factor of $0.9$,
$\lambda_{\mathrm{ana}}^{\min}=10^{-12}$, and
$\lambda_{\mathrm{ana}}^{\max}=10^{-3}$.  The legality coefficient is a
nonnegative projected dual variable for
$E_{\mathrm{leg}}\leq10^{-4}$, initialized at zero and updated by projected
Adam ascent with learning rate $5\times10^{-4}$ while $t<100$, with a maximum value of 15 and a reference floor of
$\lambda_{\mathrm{floor}}=0.1$. After the standard cell response is formed,
the macro coordinates are refined jointly for $K=10$ Adam steps with learning
rate $10^{-2}$ using the composite energy in
\Eqref{eq:combined_placement_energy}.

\subsection{Thermal Guidance}
\label{sec:thermal-guidance-details}

The frozen thermal model produces predictions for both individual macros and the whole chip,
\begin{equation}
F_{\psi}(\mathbf p,\mathbf c_{\mathrm{th}})
=
\left(\widehat{\mathbf t},\widehat{\mathbf T}\right),
\quad
\widehat{\mathbf t}=[\widehat t_1,\ldots,\widehat t_{N_M}],
\quad
\widehat{\mathbf T}=
[\widehat T_{\mathrm{mean}},\widehat T_{95},\widehat T_{\mathrm{peak}}].
\label{eq:thermal_surrogate_output}
\end{equation}
Only $\widehat{\mathbf t}$ enters the objective that combines area temperature and hotspots in
Equations~\ref{eq:thermal_macro_statistics}--\ref{eq:thermal_base_energy}. The
global head is retained for calibration diagnostics.  The reported objective
uses $\rho=0.8$ and $\tau_T=2.0\,^{\circ}\mathrm{C}$.

The thermal coefficient is controlled by a detached gradient budget.  At
inner step $k$, let
\begin{equation}
R_{\mathrm{base}}
=\sqrt{
\operatorname{RMS}(\lambda_{\mathrm{leg}}\nabla E_{\mathrm{leg}})^2
+
\operatorname{RMS}(\lambda_{\mathrm{ana}}\nabla E_{\mathrm{ana}})^2},
\qquad
R_{\mathrm{th}}=\operatorname{RMS}(\nabla E_{\mathrm{th}}).
\end{equation}
Its instantaneous target is
\begin{equation}
\lambda_{\mathrm{th},k}^{\mathrm{tar}}
=
\operatorname{clip}\!\left(
r_{\mathrm{th}}\frac{R_{\mathrm{base}}}{R_{\mathrm{th}}+\epsilon},
\lambda_{\mathrm{th}}^{\min},
\lambda_{\mathrm{th}}^{\max}
\right).
\label{eq:thermal_target_weight}
\end{equation}
The target is smoothed over inner steps and modulated by a unified gate,
\begin{equation}
\lambda_{\mathrm{th},k}(t)
=
\alpha_{\mathrm{gate},k}(t)
\lambda_{\mathrm{th},k}^{\mathrm{tar}},
\qquad
\alpha_{\mathrm{gate},k}(t)\in[0,1].
\label{eq:thermal_guidance_schedule}
\end{equation}
The gate attenuates thermal guidance when the current placement is infeasible,
when the thermal update conflicts with legality, or at the terminal diffusion step.

We use $r_{\mathrm{th}}=1.0$,
$\lambda_{\mathrm{th}}^{\min}=0$,
$\lambda_{\mathrm{th}}^{\max}=2.0$. Thermal,
analytical, and legality energies are differentiated in the same $K=10$ Adam
inner loop with learning rate $10^{-2}$. The model architecture, supervision, and
training settings are given in Appendix~\ref{sec:thermal-model-details}.

\section{Complete Energy Guidance Ablation Results}
\label{sec:complete-ablation-results}

This section complements the primary $\text{HPWL}_{\text{DP}}$ and peak temperature curves in Figure~\ref{fig:energy-guidance-ablation} with normalized routed wirelength and 95th percentile (p95) temperature results, as well as absolute means and mean ratios for each benchmark and metric.

\begin{table}[!t]
	\centering
	\caption{Summary of the four energy guidance variants over 20 benchmarks.
		Each metric is first averaged over three random seeds for every benchmark
		and variant. Each entry reports the arithmetic mean, followed in parentheses
		by the mean ratio to the unguided w/o A\&T variant across benchmarks. mHPWL,
		$\text{HPWL}_{\text{DP}}$, and Routed WL are reported in units of
		$10^{8}$, $10^{9}$, and $10^{6}\,\mu\mathrm{m}$, respectively. mHPWL and
		$\text{HPWL}_{\text{DP}}$ are proxies, whereas Routed WL is measured after ORFS routing.
		Temperature ratios are computed from the temperature rise above the
		$318.15\,\mathrm{K}$ ($45\,^{\circ}\mathrm{C}$) ambient temperature. All temperature values are
		measured by HotSpot rather than estimated by the thermal surrogate and are
		reported in $^{\circ}\mathrm{C}$.}
	\label{tab:ablation-summary}
	\begin{tabular}{@{}lccccc@{}}
		\toprule
		Variant & mHPWL & $\text{HPWL}_{\text{DP}}$ & Routed WL & Peak T & P95 T \\
		\midrule
		w/o A\&T & 1.07 (1.000) & 6.67 (1.000) & 5.47 (1.000) & 61.89 (1.000) & 61.22 (1.000) \\
		w/o A & 1.26 (1.159) & 6.93 (1.027) & 5.40 (0.998) & 61.19 (0.960) & 60.56 (0.958) \\
		w/o T & 1.53 (1.424) & 5.90 (0.922) & 4.94 (0.940) & 61.77 (0.994) & 61.12 (0.994) \\
		EGGPlacer & 1.71 (1.550) & 6.08 (0.949) & 4.99 (0.958) & 61.50 (0.980) & 60.86 (0.981) \\
		\bottomrule
	\end{tabular}
\end{table}


\begin{figure}[!t]
	\centering
	\begin{subfigure}[t]{0.49\linewidth}
		\centering
		\includegraphics[width=\linewidth]{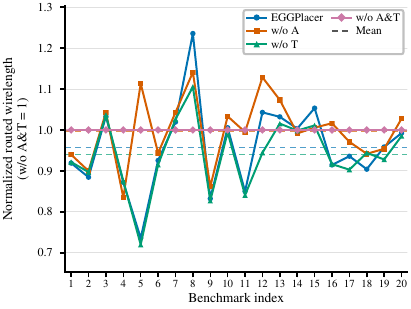}
		\caption{Routed wirelength after ORFS.}
		\label{fig:ablation-wirelength}
	\end{subfigure}
	\hfill
	\begin{subfigure}[t]{0.49\linewidth}
		\centering
		\includegraphics[width=\linewidth]{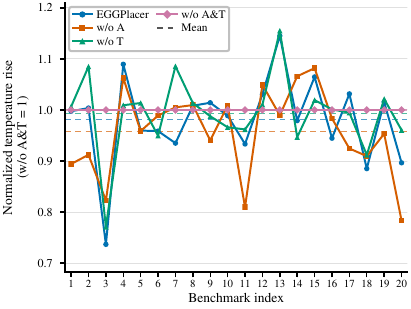}
		\caption{HotSpot p95 temperature.}
		\label{fig:ablation-p95-temperature}
	\end{subfigure}
	\caption{Additional ablation results for analytical and thermal guidance.
		Means over three seeds are normalized per benchmark to EGGPlacer w/o A\&T. Lower
		is better. Benchmark indices use the mapping defined in
		Figure~\ref{fig:energy-guidance-ablation}. P95 temperature is normalized as the rise above
		$318.15\,\mathrm{K}$ ($45\,^{\circ}\mathrm{C}$). Markers denote benchmarks,
		solid lines connect each variant's results, and dashed lines denote variant means.}
	\label{fig:energy-guidance-ablation-additional}
	\vspace{-5mm}
\end{figure}

Figure~\ref{fig:energy-guidance-ablation-additional} reports the two complementary curves for each benchmark, and Table~\ref{tab:ablation-summary} provides the complete numerical summary.
Analytical guidance (w/o T) substantially reduces both $\text{HPWL}_{\text{DP}}$ and routed wirelength despite increasing mHPWL. The complete method shows the same trend, reinforcing that mHPWL, which ignores standard cell placement, is not a reliable proxy for mixed-size placement quality or routed wirelength.

Thermal guidance alone (w/o A) yields the lowest peak and p95 temperature rises. When combined with analytical guidance, it continues to improve both thermal metrics, although the marginal benefit is smaller and $\text{HPWL}_{\text{DP}}$ increases slightly. The surrogate calibration analysis below examines whether reduced surrogate accuracy under joint guidance may help explain the smaller marginal thermal benefit.

\subsection{Correlation between Placement Proxies and Routed Wirelength}
\label{sec:metric-proxy-correlation}

We further quantify, within the EGGPlacer framework, how reliably mHPWL and $\text{HPWL}_{\text{DP}}$ predict final routed wirelength.
We restrict this analysis to the four guidance variants on all 20 benchmarks, thereby avoiding systematic differences among unrelated placement algorithms and focusing directly on the effect of enabling A and T guidance.
Directly pooling raw values across circuits is inappropriate because circuit scale simultaneously increases mHPWL, $\text{HPWL}_{\text{DP}}$, and routed wirelength, creating a high correlation even when the variants are ranked differently within each circuit.
We therefore divide each metric by its geometric mean across the four variants within the same benchmark and take the logarithm before computing correlations.

\begin{figure}[htbp]
	\centering
	\begin{subfigure}[t]{0.49\linewidth}
		\centering
		\includegraphics[width=\linewidth]{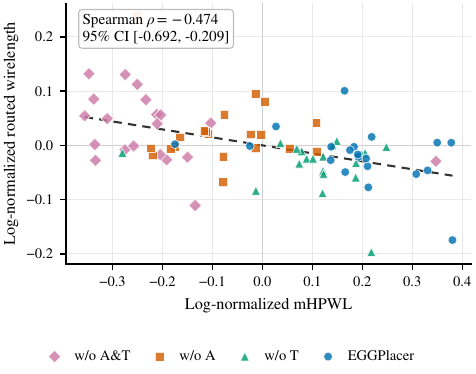}
		\caption{mHPWL vs. routed wirelength.}
		\label{fig:metric-proxy-mhpwl-routed-wirelength}
	\end{subfigure}
	\hfill
	\begin{subfigure}[t]{0.49\linewidth}
		\centering
		\includegraphics[width=\linewidth]{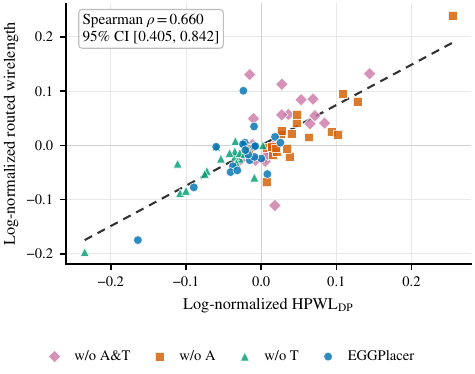}
		\caption{$\text{HPWL}_{\text{DP}}$ vs. routed wirelength.}
		\label{fig:metric-proxy-hpwl-routed-wirelength}
	\end{subfigure}
	\caption{Correlations within each design between wirelength proxies and routed wirelength.
		Each marker denotes the mean over three seeds for a variant and benchmark after log
		normalization by the geometric mean across the four EGGPlacer variants for
		that benchmark. The analysis contains four variants and all 20 benchmarks
		(80 observations). Dashed lines are pooled linear fits for visualization.
		Spearman confidence intervals are obtained from 10,000 bootstrap samples
		that resample complete benchmarks.}
	\label{fig:metric-proxy-correlations}
	\vspace{-6mm}
\end{figure}

Figure~\ref{fig:metric-proxy-correlations}(a) shows that mHPWL does not exhibit the positive monotonic association expected from a reliable proxy for routed wirelength within the EGGPlacer ablation family. Its Spearman correlation with routed wirelength is $-0.474$ (95\% bootstrap CI $[-0.692,-0.209]$), with a Pearson correlation of $-0.482$.
The mean Spearman correlation computed separately within each benchmark is also negative at $-0.420$.
This inverse association should not be interpreted causally as evidence that increasing mHPWL necessarily improves routed wirelength, nor is it claimed to hold across arbitrary placers. Instead, it demonstrates that rankings among the four guidance configurations based on mHPWL do not reliably predict routed outcomes.

In contrast, the DREAMPlace-refined proxy $\text{HPWL}_{\text{DP}}$ is more informative about routed wirelength, as shown in Figure~\ref{fig:metric-proxy-correlations}(b), although the association remains moderate rather than deterministic.
The corresponding Spearman and Pearson correlations are $0.660$ (95\% CI $[0.405,0.842]$) and $0.787$, and the mean Spearman correlation within each benchmark is $0.520$.
These results support using $\text{HPWL}_{\text{DP}}$ as an inexpensive intermediate proxy while retaining routed wirelength for downstream claims.

\subsection{Engine Dependence of Placement Proxies for Routed Wirelength}
\label{sec:hpwl-stage-consistency}

The preceding analysis uses the mixed-size placement produced by the common DREAMPlace refinement stage. We denote its HPWL proxy by $\text{HPWL}_{\text{DP}}$.
DREAMPlace builds on the electrostatic placement formulation of ePlace and RePlAce~\citep{nonlinear7eplace,nonlinear8replace,dreamplaceDAC,dreamplaceTCAD}. The ORFS placement flow uses RePlAce-based global placement followed by detailed placement and other transformations specific to the flow~\citep{kahng2021openroad}, yielding a second HPWL proxy, $\text{HPWL}_{\text{ORFS}}$, before routing. The two proxies therefore arise from closely related analytical foundations but different standard cell placement flows.

We quantify this engine dependence using the same controlled protocol as Section~\ref{sec:metric-proxy-correlation}. It includes the four EGGPlacer guidance variants on all 20 benchmarks, log normalization by the geometric mean within each design, and 10,000 bootstrap samples at the benchmark level.
In addition to correlation, we report pairwise rank agreement, namely the fraction of the six variant pairs per benchmark that receive the same ordering under both metrics.

\begin{table}[htbp]
	\centering
	\caption{Engine dependence of placement proxy fidelity to routed
		wirelength. DP and ORFS denote $\text{HPWL}_{\text{DP}}$ and
		$\text{HPWL}_{\text{ORFS}}$, respectively, and RWL denotes final
		routed wirelength. Correlations use 80 observations, each averaged over
		three seeds and normalized within its design. Confidence intervals resample complete benchmarks.}
	\label{tab:hpwl-stage-consistency}
	\begin{tabular}{@{}lcccc@{}}
		\toprule
		Relation & Pearson $r$ & Spearman $\rho$ (95\% CI) & Mean design $\rho$ & Rank agreement \\
		\midrule
		DP vs. ORFS  & 0.861 & 0.684 $[0.433,0.856]$ & 0.560 & 74.2\% \\
		DP vs. RWL   & 0.787 & 0.660 $[0.405,0.842]$ & 0.520 & 72.5\% \\
		ORFS vs. RWL & 0.941 & 0.918 $[0.817,0.977]$ & 0.880 & 91.7\% \\
		\bottomrule
	\end{tabular}
\end{table}

Table~\ref{tab:hpwl-stage-consistency} shows that the two HPWL values are positively associated but still disagree on 31 of the 120 pairwise orderings within designs. They also select the same best variant on only 8 of the 20 benchmarks. Thus, closely related analytical formulations do not guarantee equivalent rankings when used in different standard cell placement flows.
The association with routed wirelength is substantially stronger for $\text{HPWL}_{\text{ORFS}}$ than for $\text{HPWL}_{\text{DP}}$. Spearman correlation increases from $0.660$ to $0.918$, pairwise rank agreement increases from $72.5\%$ to $91.7\%$, and the number of benchmarks with the same best variant increases from 8 to 18.
This stronger association is consistent with target flow alignment because $\text{HPWL}_{\text{ORFS}}$ is measured within the same downstream flow that produces the routed wirelength.

The distinction also explains an otherwise counterintuitive comparison between the standalone DREAMPlace baseline and EGGPlacer.
Across the 20 benchmarks with complete data for both methods, DREAMPlace has lower $\text{HPWL}_{\text{DP}}$ on 17 designs, but lower $\text{HPWL}_{\text{ORFS}}$ on only 7 and lower routed wirelength on 9.
Relative to the final routed ordering, their ordering is reversed on 8 benchmarks when using $\text{HPWL}_{\text{DP}}$, compared with only 4 when using $\text{HPWL}_{\text{ORFS}}$.
Thus, a macro configuration can receive different rankings under different standard cell placement flows and the final routed evaluation.
This result reinforces the paper's use of routed wirelength for downstream claims and treats HPWL as a diagnostic whose fidelity can vary across placement engines.

\section{Thermal Model}
\label{sec:thermal-model-details}

\subsection{Training Data and Targets}
\label{sec:thermal-training-data}

The thermal dataset contains 1,200 placements simulated by HotSpot from all 20
benchmarks. The 200 base placements were generated by DREAMPlace and earlier
versions of EGGPlacer. We reuse these placements without a separate data
construction procedure tailored to the thermal surrogate. We generate five additional
samples from each base placement by perturbing macro coordinates with Gaussian
offsets at 1--5\% of the die dimensions, giving six samples per base placement.
Only macro locations are changed. The design geometry and power context remain
fixed.

We assign each base placement and its five perturbations to the same data
partition. The resulting training, validation, and test partitions contain
960, 120, and 120 placements, respectively.  All normalization statistics are
estimated from the training partition only.

For thermal evaluation, we employ a layered thermal model comprising an active silicon die, a thermal interface material (TIM), an aluminum heat spreader, and an aluminum heat sink. The thermal conductivities of silicon, TIM, and aluminum are set to $130$, $4$, and $237$ $\mathrm{W/(m\cdot K)}$, respectively. Heat dissipation from the heat sink to the ambient environment is modeled using an equivalent convective thermal resistance of $0.1$ $\mathrm{K/W}$. Since most macros in ChiPBench
are SRAM blocks with relatively moderate thermal effects, we scale the original
macro power values by a factor of 10 to approximate the thermal behavior of
computational macros with high power consumption. This amplification produces
larger spatial temperature gradients and more pronounced hotspots, thereby
providing a controlled stress test for evaluating whether the placer can effectively
respond to and mitigate severe thermal constraints.

OpenSTA~\citep{opensta} supplies a fixed vectorless power context for each design, and HotSpot
produces the corresponding temperature field at steady state on a $64\times64$
grid. The predictor is supervised by one temperature averaged over the area of each macro
and three statistics for the whole chip, namely mean, 95th percentile, and peak temperature.
Holding the power context fixed across perturbations encourages the surrogate
to learn thermal changes induced by placement rather than power reassignment.
The surrogate is used only to provide differentiable thermal guidance signals
during sampling, while all reported thermal results are independently
recomputed with HotSpot.

\subsection{Network Architecture}
\label{sec:thermal-predictor-architecture}

The thermal predictor is a Transformer~\citep{Ashish2017Attention} that directly processes a set of macros.
Each macro location is represented by continuous coordinates, eliminating the need to rasterize the placement. For macro
$i$, its center $\mathbf{p}_i=(x_i,y_i)$ normalized by the die dimensions is encoded using four
Fourier bands,
\begin{equation}
\boldsymbol{\phi}(\mathbf{p}_i)
=\left[\mathbf{p}_i,
\left\{\sin(\omega_k\mathbf{p}_i),
\cos(\omega_k\mathbf{p}_i)\right\}_{k=0}^{3}\right],
\qquad \omega_k=\pi 2^k,
\label{eq:thermal-fourier-encoding}
\end{equation}
which gives a differentiable coordinate representation with 18 dimensions. This
representation is concatenated with five static macro features (width, height,
area, power, and power density) and
an embedding of dimension 32 for the macro type. The resulting vector has 55 dimensions
and is projected to a token with 128 dimensions by a linear layer, layer
normalization, and GELU activation.

A learnable global token is conditioned on two design features:
the total macro power and macro count.  The global and macro
tokens are processed by three Transformer encoder layers with normalization before each layer. Each layer
has eight attention heads, a feedforward block of dimension 512, GELU
activation, and dropout of 0.1. Separate MLP heads with two layers predict one
temperature for every macro token and the three temperatures for the whole chip from
the global token.  Padding masks accommodate different macro counts.
Because no embedding of the sequence index is used, the macro outputs are
equivariant to permutation and the outputs for the whole chip are invariant to permutation.
Physical location is represented solely by continuous Fourier features,
preserving the differentiability with respect to macro coordinates needed for
thermal guidance. Overall, the thermal predictor has 655,684
trainable parameters, and its macro type embedding contains 66 entries.

\subsection{Training Objective}
\label{sec:thermal-predictor-loss}

Let $h_{\delta}(e)$ denote the Huber loss with $\delta=0.5\,^{\circ}\mathrm{C}$,
and let $v_{b,i}$ indicate a valid macro that is not padding. The term for each macro is
\begin{equation}
\mathcal{L}_{\mathrm{macro}}
=\frac{\sum_{b,i}v_{b,i}
h_{\delta}(\widehat{T}_{b,i}-T_{b,i})}
{\sum_{b,i}v_{b,i}}.
\label{eq:thermal-macro-loss}
\end{equation}
For $q\in\{\mathrm{mean},\mathrm{p95},\mathrm{peak}\}$, define
$\mathcal{L}_{q}=B^{-1}\sum_b
h_{\delta}(\widehat{T}_{b,q}-T_{b,q})$. The loss for the whole chip emphasizes tail
temperature while retaining supervision of the other statistics:
\begin{equation}
\mathcal{L}_{\mathrm{global}}
=\frac{\lambda_1\mathcal{L}_{\mathrm{mean}}
+\lambda_2\mathcal{L}_{\mathrm{p95}}
+\lambda_3\mathcal{L}_{\mathrm{peak}}}
{\lambda_1+\lambda_2+\lambda_3}.
\label{eq:thermal-global-loss}
\end{equation}
We set $\lambda_1=0.2$, $\lambda_2=0.4$, and $\lambda_3=1.0$.
We additionally penalize physically inconsistent ordering of the three
predictions,
\begin{equation}
\mathcal{L}_{\mathrm{ord}}
=\frac{1}{B}\sum_b\left[
\operatorname{ReLU}(\widehat{T}_{b,\mathrm{mean}}
-\widehat{T}_{b,\mathrm{p95}})
+\operatorname{ReLU}(\widehat{T}_{b,\mathrm{p95}}
-\widehat{T}_{b,\mathrm{peak}})\right].
\label{eq:thermal-ordering-loss}
\end{equation}
The complete objective is
\begin{equation}
\mathcal{L}_{\mathrm{thermal}}
=\mathcal{L}_{\mathrm{global}}
+\mathcal{L}_{\mathrm{macro}}+0.1\mathcal{L}_{\mathrm{ord}}.
\label{eq:thermal-training-objective}
\end{equation}
Dense supervision for each macro encourages the attention layers to capture local
thermal interactions and directly supports the guidance objective.  The
weighted global term instead supports diagnostic calibration against the
statistics for the whole chip used in downstream evaluation.

\subsection{Optimization Settings}
\label{sec:thermal-predictor-training}

We train the predictor with AdamW using batch size 8, an initial learning rate
of $2\times10^{-4}$, and weight decay $10^{-4}$.  The learning rate is reduced
on a validation plateau, gradients are clipped to an $\ell_2$ norm of 1.0,
and early stopping selects the checkpoint with the lowest validation loss. Random
horizontal and vertical flips provide augmentation that preserves placement properties.

\subsection{Surrogate Calibration under Guided Sampling}
\label{sec:thermal-surrogate-calibration}

We further examine whether optimizing the thermal surrogate changes its agreement with the downstream HotSpot evaluation.
This analysis reuses the predictor and HotSpot temperatures of the four ablation variants on all 20 benchmarks.

\begin{figure}[H]
	\centering
	\begin{subfigure}[t]{0.49\linewidth}
		\centering
		\includegraphics[width=\linewidth]{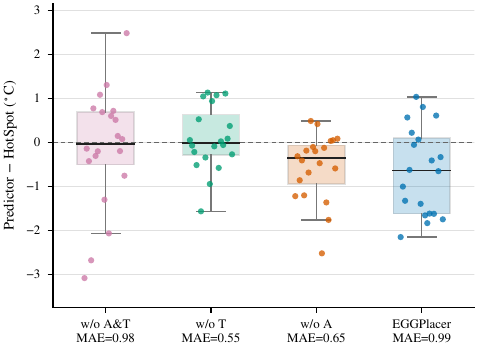}
		\caption{Peak temperature prediction error.}
		\label{fig:thermal-surrogate-peak-error}
	\end{subfigure}
	\hfill
	\begin{subfigure}[t]{0.49\linewidth}
		\centering
		\includegraphics[width=\linewidth]{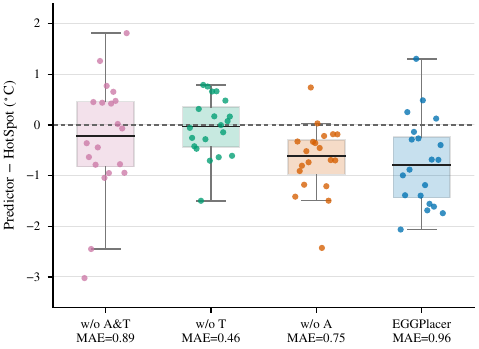}
		\caption{P95 temperature prediction error.}
		\label{fig:thermal-surrogate-p95-error}
	\end{subfigure}

	\begin{subfigure}[t]{0.49\linewidth}
		\centering
		\includegraphics[width=\linewidth]{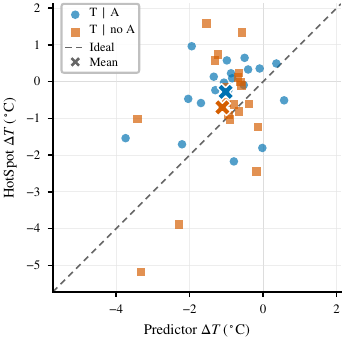}
		\caption{Effect of enabling T on peak temperature.}
		\label{fig:thermal-surrogate-peak-guidance}
	\end{subfigure}
	\hfill
	\begin{subfigure}[t]{0.49\linewidth}
		\centering
		\includegraphics[width=\linewidth]{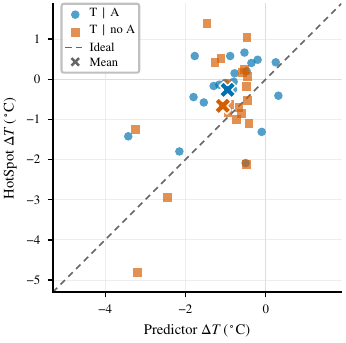}
		\caption{Effect of enabling T on p95 temperature.}
		\label{fig:thermal-surrogate-p95-guidance}
	\end{subfigure}
	\caption{Calibration of the thermal surrogate on the four EGGPlacer ablation
		variants. Panels (a) and (b) show signed errors averaged over three seeds for
		individual benchmarks. Panels (c) and (d) compare temperature changes predicted
		by the surrogate with those measured by HotSpot after enabling thermal guidance. Here,
		$\Delta T=T_{\mathrm{enabled}}-T_{\mathrm{disabled}}$, where $T_{\mathrm{enabled}}$
		and $T_{\mathrm{disabled}}$ are temperatures from otherwise identical variants with
		thermal guidance enabled and disabled, respectively. Negative values indicate improvement.
		``T $\mid$ A'' and ``T $\mid$ no A'' denote comparisons with and without analytical
		guidance, respectively. Crosses denote
		means, and the dashed diagonal denotes ideal agreement.}
	\label{fig:thermal-surrogate-bias}
\end{figure}

Figures~\ref{fig:thermal-surrogate-bias}(a) and~\ref{fig:thermal-surrogate-bias}(b) report the signed error, defined as the predictor temperature minus the HotSpot temperature.
When analytical guidance is present, enabling thermal guidance increases the mean absolute error for both peak and 95th percentile temperatures, with larger errors on 14 of the 20 benchmarks. In contrast, thermal guidance reduces both errors when analytical guidance is absent, indicating that the degradation is specific to the joint guidance setting.

Figures~\ref{fig:thermal-surrogate-bias}(c) and~\ref{fig:thermal-surrogate-bias}(d) compare the temperature improvement predicted by the surrogate with that measured by HotSpot. Under joint guidance, the surrogate predicts an average peak temperature reduction of $1.020\,^{\circ}\mathrm{C}$, whereas HotSpot measures $0.278\,^{\circ}\mathrm{C}$. A similar discrepancy appears for the 95th percentile temperature, while the peak temperature gap is smaller without analytical guidance.

These results suggest that joint analytical and thermal optimization may induce a distribution shift under which the surrogate overestimates the marginal improvement in tail temperature.
Indeed, the HotSpot results in the main paper still confirm that the complete method improves thermal quality over the unguided baseline.
A promising direction is to augment surrogate training with placements generated by joint guidance, emphasize examples with large errors in tail temperature, and use guidance informed by uncertainty or constrained by a trust region to prevent optimization from moving too far beyond the region where the surrogate is well calibrated.

\clearpage
\section{Example Visualizations of Placement and Thermal Simulation Results}
\label{sec:qualitative-placement-thermal-results}

\begin{figure}[!htbp]
	\centering
	\includegraphics[width=0.95\linewidth]{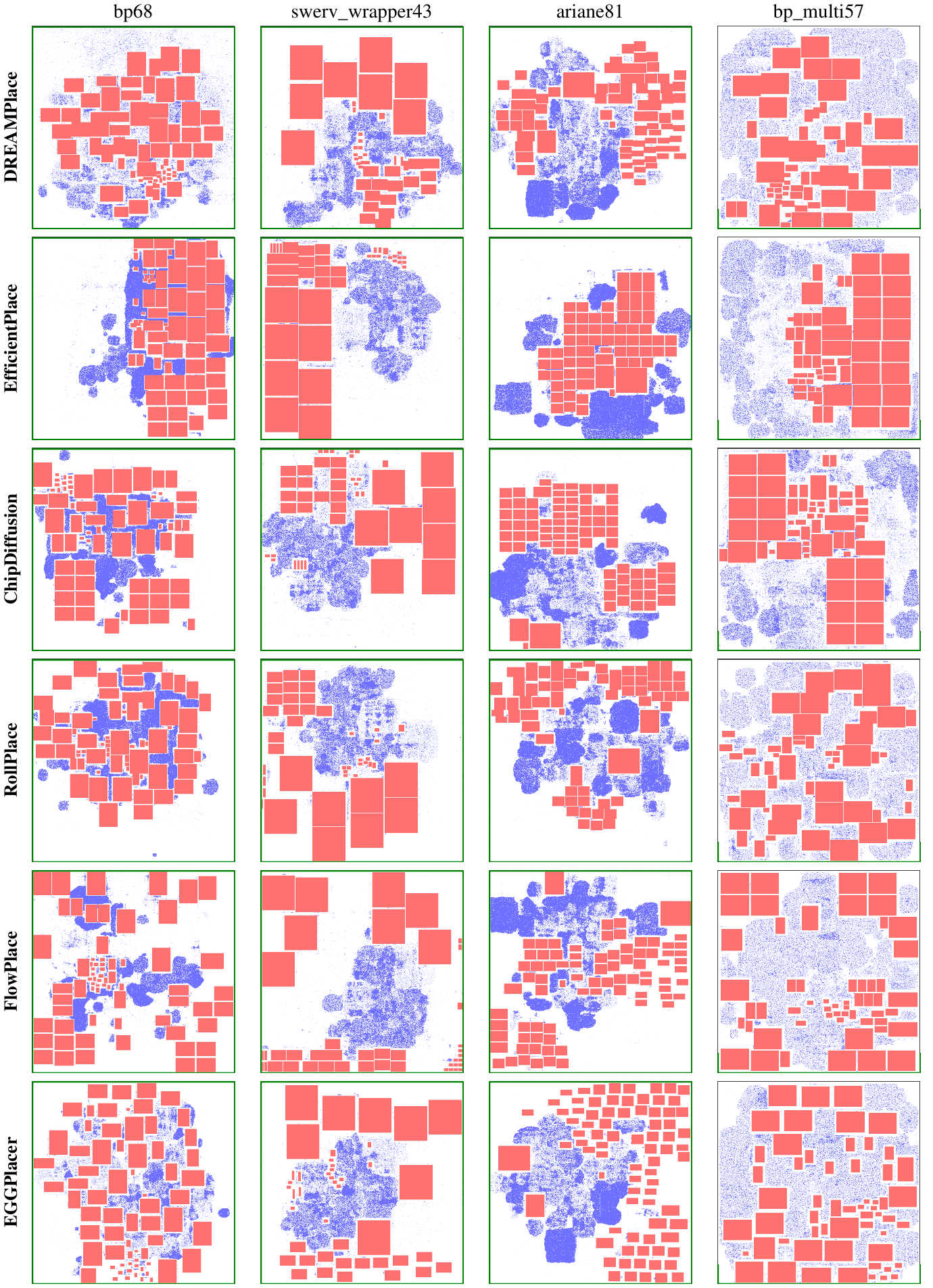}
	\caption{ORFS global placement results obtained with macro locations generated
		by six placement methods on four representative designs. Red rectangles
		denote macros, blue points denote standard cells, and green points denote
		terminals.}
	\label{fig:global-placement-visualization}
\end{figure}

\begin{figure}[!htbp]
	\centering
	\includegraphics[width=\linewidth]{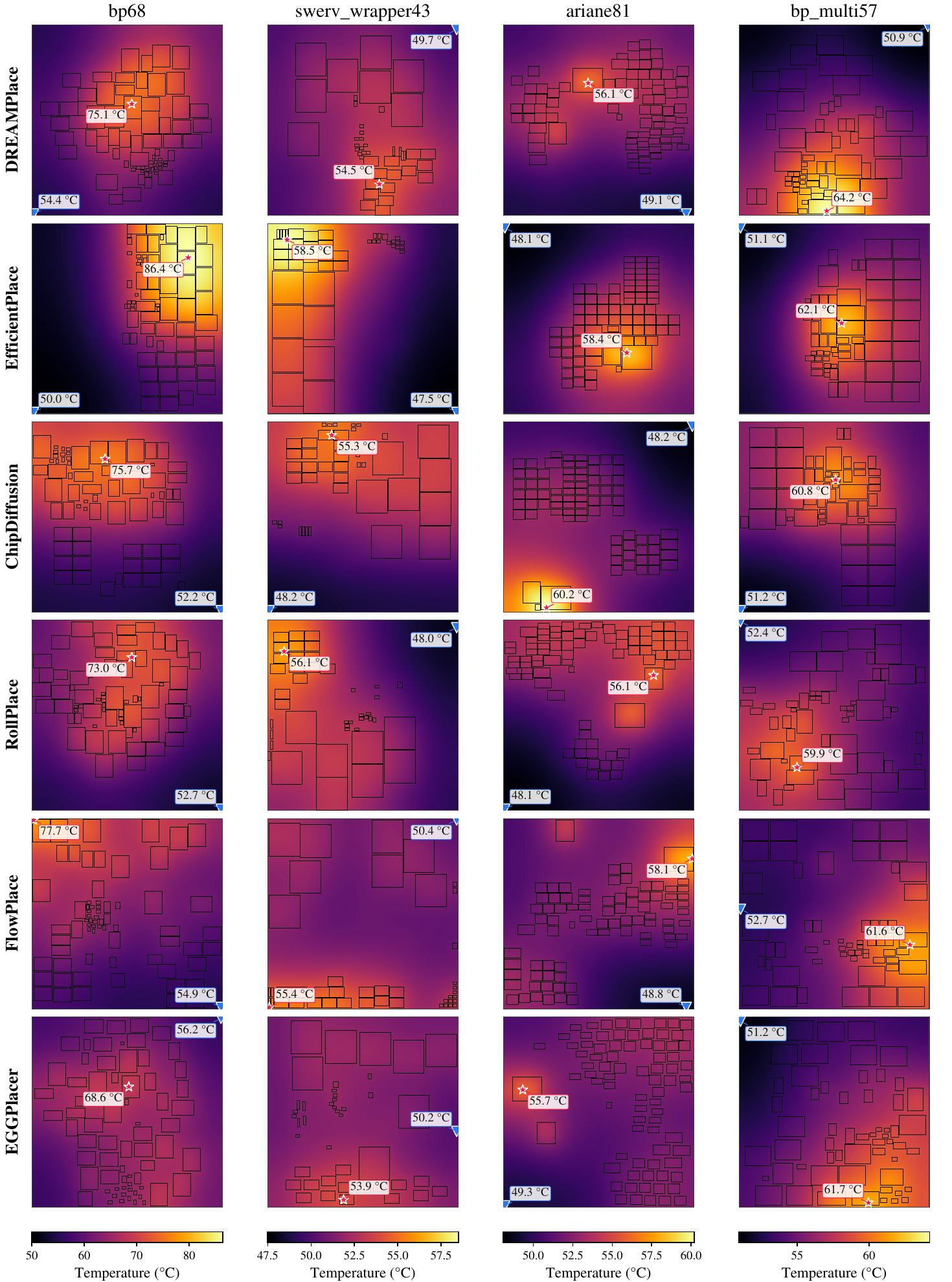}
	\caption{HotSpot temperature distributions at steady state for the placements
		in Figure~\ref{fig:global-placement-visualization}. Red stars and blue
		triangles mark the hottest and coolest locations, respectively. Each design
		uses a common temperature scale across methods, shown below its column.}
	\label{fig:thermal-simulation-visualization}
\end{figure}

Figure~\ref{fig:global-placement-visualization} compares the ORFS global
placements obtained with macro locations generated by the six evaluated
methods on four representative designs. Figure~\ref{fig:thermal-simulation-visualization}
shows the corresponding HotSpot temperature distributions at steady state.

\FloatBarrier

\clearpage
\section{Complete Baseline Results for Each Design}
\label{sec:complete-baseline-results}
\thispagestyle{fancy}

Tables~\ref{tab:mhpwl_benchmark_comparison_full}--%
\ref{tab:openroad_benchmark_comparison_full} provide the corresponding mixed-size placement and end-to-end results for each design.


\begin{table}[H]
	\centering
	\caption{mHPWL results for each design ($10^{8}$). sw and sw43 abbreviate
		swerv\_wrapper and swerv\_wrapper43. EGGPlacer reports means over three seeds
		$\pm$ sample standard deviations. Bold and underlined values rank first and
		second.}
	\label{tab:mhpwl_benchmark_comparison_full}
	\renewcommand{\arraystretch}{0.92}
	\begin{tabular}{lcccccc}
		\toprule
		Benchmark & DP & EffPlace & ChipDiff & RollPlace & FlowPlace & EGGPlacer \\
		\midrule
		ariane133        & 2.98 & \underline{2.53} & \textbf{2.18} & 3.90 & 5.37 & 3.98 $\pm$ 0.09 \\
		ariane136        & 3.95 & \underline{2.56} & \textbf{2.27} & 5.00 & 5.51 & 4.50 $\pm$ 0.31 \\
		ariane81         & 3.38 & \underline{2.47} & \textbf{2.17} & 4.99 & 4.90 & 5.35 $\pm$ 0.34 \\
		bp               & 0.48 & \textbf{0.32} & \underline{0.35} & 0.81 & 0.47 & 0.72 $\pm$ 0.16 \\
		bp68             & 0.88 & \underline{0.59} & \textbf{0.58} & 1.63 & 0.86 & 1.42 $\pm$ 0.25 \\
		bp\_be           & 0.35 & \textbf{0.21} & 0.27 & 0.28 & \underline{0.24} & 0.42 $\pm$ 0.04 \\
		bp\_be12         & 0.32 & \textbf{0.18} & \underline{0.19} & 0.41 & 0.27 & 0.40 $\pm$ 0.05 \\
		bp\_fe           & 0.21 & \textbf{0.13} & \underline{0.15} & 0.24 & 0.19 & 0.24 $\pm$ 0.01 \\
		bp\_fe38         & 1.24 & \underline{0.81} & \textbf{0.80} & 1.94 & 2.19 & 1.75 $\pm$ 0.04 \\
		bp\_multi        & 0.53 & \textbf{0.34} & \underline{0.40} & 1.19 & 0.59 & 0.71 $\pm$ 0.08 \\
		bp\_multi57      & 0.81 & \textbf{0.43} & \underline{0.45} & 1.23 & 0.88 & 1.04 $\pm$ 0.24 \\
		dft68            & 3.56 & \underline{3.54} & \textbf{2.64} & 3.67 & 3.88 & 4.45 $\pm$ 0.38 \\
		ethernet         & 0.17 & \underline{0.10} & \textbf{0.09} & 0.10 & 0.12 & 0.18 $\pm$ 0.03 \\
		isa\_npu         & 3.72 & \underline{0.48} & \textbf{0.37} & 1.04 & 0.54 & 0.96 $\pm$ 0.13 \\
		mor1kx           & 0.87 & \underline{0.61} & \textbf{0.54} & 0.76 & 0.74 & 0.79 $\pm$ 0.05 \\
		or1200           & 0.98 & \textbf{0.63} & -- & 1.19 & \underline{0.88} & 1.02 $\pm$ 0.04 \\
		sw               & 1.57 & \underline{1.44} & \textbf{1.12} & 1.83 & 2.10 & 2.24 $\pm$ 0.18 \\
		sw43             & 2.05 & \underline{1.68} & \textbf{1.33} & 3.12 & 2.48 & 2.88 $\pm$ 0.43 \\
		vga\_lcd         & 1.17 & \textbf{0.92} & \underline{1.08} & 1.44 & 1.34 & 1.18 $\pm$ 0.06 \\
		VeriGPU          & 0.04 & \textbf{0.03} & 0.03 & 0.05 & \underline{0.03} & 0.05 $\pm$ 0.005 \\
		\midrule
		Average      & 1.46 & \underline{1.00} & \textbf{0.90} & 1.74 & 1.68 & 1.71 $\pm$ 0.02 \\
		Average Rank & 3.75 & \underline{1.55} & \textbf{1.53} & 5.10 & 3.85 & 5.00 \\
		\bottomrule
	\end{tabular}
\end{table}

\begin{table}[H]
	\centering
	\caption{$\text{HPWL}_{\text{DP}}$ results for each design after DREAMPlace refinement
		($10^{9}$). sw and sw43 abbreviate swerv\_wrapper and swerv\_wrapper43.
		Other conventions follow
		Table~\ref{tab:mhpwl_benchmark_comparison_full}.}
	\label{tab:DP_benchmark_comparison_full}
	\renewcommand{\arraystretch}{0.92}
	\begin{tabular}{lcccccc}
		\toprule
		Benchmark & DP & EffPlace & ChipDiff & RollPlace & FlowPlace & EGGPlacer \\
		\midrule
		ariane133        & \textbf{6.85} & 7.55 & \underline{7.13} & 7.49 & 7.98 & 7.36 $\pm$ 0.08 \\
		ariane136        & \textbf{7.46} & 7.91 & 8.23 & 7.84 & 8.15 & \underline{7.81} $\pm$ 0.10 \\
		ariane81         & \textbf{7.18} & 8.62 & 7.86 & \underline{7.77} & 8.08 & 7.87 $\pm$ 0.10 \\
		bp               & \textbf{9.71} & 11.62 & 12.25 & \underline{10.72} & 10.83 & 11.00 $\pm$ 0.14 \\
		bp68             & \textbf{16.61} & 26.85 & 26.04 & \underline{19.60} & 30.27 & 20.51 $\pm$ 1.13 \\
		bp\_be           & \textbf{4.60} & 4.72 & 4.91 & \underline{4.72} & 4.80 & 4.89 $\pm$ 0.04 \\
		bp\_be12         & \textbf{5.71} & 6.42 & 6.52 & 6.02 & 6.16 & \underline{5.86} $\pm$ 0.03 \\
		bp\_fe           & \textbf{3.01} & 3.37 & 3.20 & \underline{3.01} & 3.19 & 3.02 $\pm$ 0.03 \\
		bp\_fe38         & 5.06 & \underline{4.62} & \textbf{4.61} & 4.67 & 5.07 & 4.68 $\pm$ 0.04 \\
		bp\_multi        & \textbf{5.46} & 7.32 & 6.85 & 6.20 & 6.39 & \underline{5.83} $\pm$ 0.04 \\
		bp\_multi57      & \textbf{6.42} & 10.06 & 10.22 & 7.90 & 9.50 & \underline{7.43} $\pm$ 0.22 \\
		dft68            & \textbf{1.95} & 2.25 & \underline{2.07} & 2.10 & 2.52 & 2.33 $\pm$ 0.16 \\
		ethernet         & \textbf{1.38} & 1.42 & 1.47 & \underline{1.39} & 1.39 & 1.39 $\pm$ 0.01 \\
		isa\_npu         & 12.33 & \underline{11.28} & 11.41 & \textbf{10.96} & 11.67 & 11.62 $\pm$ 0.50 \\
		mor1kx           & \textbf{2.85} & 2.99 & 2.96 & \underline{2.92} & 3.17 & 2.96 $\pm$ 0.04 \\
		or1200           & 1.70 & 1.82 & -- & 1.52 & \underline{1.51} & \textbf{1.43} $\pm$ 0.03 \\
		sw               & \underline{4.71} & 4.80 & \textbf{4.63} & 5.84 & 4.92 & 5.27 $\pm$ 0.09 \\
		sw43             & \textbf{6.11} & 6.39 & 6.52 & 6.33 & 6.75 & \underline{6.33} $\pm$ 0.14 \\
		vga\_lcd         & \textbf{1.95} & \underline{2.08} & 2.10 & 2.15 & 2.19 & 2.28 $\pm$ 0.08 \\
		VeriGPU          & \textbf{1.77} & \underline{1.79} & 1.83 & 1.91 & 1.80 & 1.80 $\pm$ 0.04 \\
		\midrule
		Average      & \textbf{5.64} & 6.69 & 6.89 & \underline{6.05} & 6.82 & 6.08 $\pm$ 0.08 \\
		Average Rank & \textbf{1.65} & 4.20 & 4.16 & \underline{2.95} & 4.65 & 3.30 \\
		\bottomrule
	\end{tabular}
\end{table}


\begin{table}[H]
	\centering
	\caption{Routed wirelength results for each design ($10^{6}\,\mu\mathrm{m}$). sw and sw43
		abbreviate swerv\_wrapper and swerv\_wrapper43. EGGPlacer reports means over
		three seeds $\pm$ sample standard deviations. Baselines are single runs. Bold and
		underlined values rank first and second.}
	\label{tab:routed_wirelength_benchmark_comparison_full}
	\renewcommand{\arraystretch}{0.92}
	\begin{tabular}{lcccccc}
		\toprule
		Benchmark & DP & EffPlace & ChipDiff & RollPlace & FlowPlace & EGGPlacer \\
		\midrule
		ariane133        & \textbf{7.02} & 7.54 & 7.39 & 7.73 & 7.54 & \underline{7.29} $\pm$ 0.27 \\
		ariane136        & \textbf{7.14} & 7.79 & 7.89 & 7.39 & 8.04 & \underline{7.30} $\pm$ 0.14 \\
		ariane81         & 6.64 & 8.42 & 7.39 & \underline{6.47} & \textbf{6.45} & 6.66 $\pm$ 0.40 \\
		bp               & 10.54 & 11.94 & 10.18 & \underline{9.59} & \textbf{9.39} & 10.32 $\pm$ 0.43 \\
		bp68             & 20.09 & 22.32 & 16.79 & \underline{13.94} & 19.09 & \textbf{12.97} $\pm$ 0.58 \\
		bp\_be           & 4.24 & \underline{4.19} & 4.77 & \textbf{4.00} & 4.32 & 4.32 $\pm$ 0.23 \\
		bp\_be12         & \underline{4.13} & 4.18 & 4.59 & \textbf{4.01} & 4.21 & 4.13 $\pm$ 0.03 \\
		bp\_fe           & 3.44 & 3.60 & 3.36 & \textbf{3.04} & \underline{3.22} & 3.48 $\pm$ 0.06 \\
		bp\_fe38         & 3.47 & 3.72 & 4.85 & 3.54 & \underline{3.17} & \textbf{3.00} $\pm$ 0.04 \\
		bp\_multi        & 5.71 & 5.98 & 5.88 & 6.04 & \textbf{5.41} & \underline{5.65} $\pm$ 0.11 \\
		bp\_multi57      & 7.39 & 8.61 & 9.71 & 6.76 & \underline{6.68} & \textbf{6.38} $\pm$ 0.34 \\
		dft68            & \textbf{1.25} & 1.45 & \underline{1.30} & 1.30 & 1.47 & 1.36 $\pm$ 0.08 \\
		ethernet         & 1.14 & \underline{1.10} & 1.23 & 1.14 & \textbf{1.09} & 1.13 $\pm$ 0.02 \\
		isa\_npu         & 7.91 & 8.16 & \underline{7.77} & 7.80 & \textbf{7.75} & 7.81 $\pm$ 0.05 \\
		mor1kx           & 2.44 & \underline{2.43} & 2.50 & \textbf{2.34} & 2.49 & 2.67 $\pm$ 0.11 \\
		or1200           & 1.04 & 1.17 & -- & 1.15 & \underline{0.97} & \textbf{0.94} $\pm$ 0.03 \\
		sw               & 6.40 & \underline{4.96} & 5.51 & 5.48 & \textbf{4.69} & 5.05 $\pm$ 0.06 \\
		sw43             & 5.31 & 5.15 & 5.64 & \textbf{5.03} & 5.04 & \underline{5.04} $\pm$ 0.12 \\
		vga\_lcd         & \underline{2.58} & 2.85 & 2.61 & 3.50 & \textbf{2.52} & 2.59 $\pm$ 0.12 \\
		VeriGPU          & \underline{1.65} & 1.67 & 1.69 & 1.67 & 1.69 & \textbf{1.61} $\pm$ 0.03 \\
		\midrule
		Average      & 5.48 & 5.86 & 5.85 & \underline{5.10} & 5.26 & \textbf{4.99} $\pm$ 0.07 \\
		Average Rank & 3.25 & 4.40 & 4.53 & 3.10 & \underline{2.85} & \textbf{2.80} \\
		\bottomrule
	\end{tabular}
\end{table}

\begin{table}[H]
	\centering
	\caption{HotSpot peak temperature results for each design ($^{\circ}\mathrm{C}$).
		sw and sw43 abbreviate swerv\_wrapper and swerv\_wrapper43. Other conventions follow
		Table~\ref{tab:routed_wirelength_benchmark_comparison_full}.}
	\label{tab:openroad_benchmark_comparison_full}
	\renewcommand{\arraystretch}{0.92}
	\begin{tabular}{lcccccc}
		\toprule
		Benchmark & DP & EffPlace & ChipDiff & RollPlace & FlowPlace & EGGPlacer \\
		\midrule
		ariane133        & 54.28 & 54.50 & \underline{52.32} & 52.53 & 53.74 & \textbf{51.99} $\pm$ 0.14 \\
		ariane136        & 54.54 & 53.92 & \underline{52.62} & 54.96 & 53.74 & \textbf{52.39} $\pm$ 0.60 \\
		ariane81         & 56.11 & 58.44 & 60.19 & \underline{56.07} & 58.08 & \textbf{55.53} $\pm$ 0.34 \\
		bp               & 55.38 & 56.92 & 54.82 & \textbf{53.75} & 55.08 & \underline{54.40} $\pm$ 0.45 \\
		bp68             & 75.07 & 86.41 & 75.74 & \underline{73.00} & 77.66 & \textbf{69.45} $\pm$ 0.75 \\
		bp\_be           & 63.44 & 63.55 & 61.88 & 62.15 & \underline{61.62} & \textbf{60.95} $\pm$ 0.80 \\
		bp\_be12         & 60.58 & 65.50 & \underline{59.66} & 66.06 & 61.65 & \textbf{58.61} $\pm$ 0.69 \\
		bp\_fe           & 73.70 & 72.94 & 72.22 & 71.30 & \textbf{70.74} & \underline{70.95} $\pm$ 0.44 \\
		bp\_fe38         & \textbf{63.61} & 69.37 & 64.13 & 65.19 & \underline{63.68} & 64.44 $\pm$ 0.42 \\
		bp\_multi        & 65.39 & 71.32 & 66.00 & \underline{64.59} & 68.02 & \textbf{63.87} $\pm$ 0.92 \\
		bp\_multi57      & 64.21 & 62.07 & \underline{60.81} & \textbf{59.86} & 61.57 & 60.94 $\pm$ 1.64 \\
		dft68            & 76.39 & 77.56 & \underline{75.54} & 79.98 & 79.15 & \textbf{74.85} $\pm$ 0.26 \\
		ethernet         & 55.26 & 56.73 & \textbf{55.08} & \underline{55.16} & 55.64 & 55.87 $\pm$ 0.29 \\
		isa\_npu         & 51.85 & 53.50 & 52.55 & \underline{51.78} & 53.11 & \textbf{51.62} $\pm$ 0.26 \\
		mor1kx           & 59.39 & 61.02 & \textbf{58.82} & 59.65 & 59.94 & \underline{58.85} $\pm$ 0.55 \\
		or1200           & \textbf{73.98} & 83.57 & -- & 80.19 & 87.12 & \underline{75.73} $\pm$ 0.44 \\
		sw               & \underline{57.20} & 57.82 & 57.93 & \underline{57.20} & 57.36 & \textbf{56.74} $\pm$ 0.59 \\
		sw43             & \underline{54.53} & 58.50 & 55.30 & 56.12 & 55.41 & \textbf{53.81} $\pm$ 0.13 \\
		vga\_lcd         & 74.57 & 77.85 & 84.57 & 76.60 & \underline{74.05} & \textbf{73.31} $\pm$ 0.90 \\
		VeriGPU          & 69.21 & 66.12 & \textbf{64.42} & 67.39 & \underline{65.25} & 65.60 $\pm$ 2.44 \\
		\midrule
		Average      & 62.93 & 65.38 & \underline{62.35} & 63.18 & 63.63 & \textbf{61.50} $\pm$ 0.07 \\
		Average Rank & 3.55 & 5.30 & \underline{3.11} & 3.40 & 3.70 & \textbf{1.75} \\
		\bottomrule
	\end{tabular}
\end{table}


\clearpage

\section{Detailed Experiment Metrics}
\thispagestyle{fancy}

Table~\ref{tab:ablation-by-design} reports the unnormalized results underlying the summary and ablation curves for the four variants.


{
\setlength{\LTpre}{6pt}
\setlength{\LTpost}{6pt}
\renewcommand{\arraystretch}{1.12}
\begin{longtable}{
	@{\hspace{2pt}}l@{\hspace{4pt}}l@{\hspace{8pt}}
	*{5}{>{\centering\arraybackslash}p{0.49in}@{\hspace{8pt}}}
	>{\centering\arraybackslash}p{0.49in}@{\hspace{2pt}}
}
	\caption{Results for the four energy guidance variants on each design,
		averaged over three seeds. mHPWL and $\text{HPWL}_{\text{DP}}$ are
		reported in units of $10^{8}$ and $10^{9}$, respectively, while
		$\text{HPWL}_{\text{ORFS}}$ and routed wirelength are reported in units
		of $10^{6}\,\mu\mathrm{m}$. Peak and p95 temperatures are measured by HotSpot and reported
		in $^{\circ}\mathrm{C}$. sw and sw43 denote swerv\_wrapper and
		swerv\_wrapper43.}
	\label{tab:ablation-by-design} \\
	\toprule
	Design & Variant & mHPWL & \shortstack{$\text{HPWL}$ \\ $\text{DP}$}
	& \shortstack{$\text{HPWL}$ \\ $\text{ORFS}$}
	& \shortstack{Routed \\ WL} & Peak T & P95 T \\
	\midrule
	\endfirsthead
	\caption[]{Results for each design for the four energy guidance variants
		(continued).} \\
	\toprule
	Design & Variant & mHPWL & \shortstack{$\text{HPWL}$ \\ $\text{DP}$}
	& \shortstack{$\text{HPWL}$ \\ $\text{ORFS}$}
	& \shortstack{Routed \\ WL} & Peak T & P95 T \\
	\midrule
	\endhead
	\midrule
	\multicolumn{8}{r}{Continued on next page} \\
	\endfoot
	\bottomrule
	\endlastfoot
	\multirow{4}{*}{ariane133} & w/o A\&T & 2.807 & 7.750 & 6.732 & 7.936 & 51.88 & 51.65 \\*
	 & w/o A & 3.424 & 7.532 & 6.291 & 7.458 & 51.27 & 50.94 \\*
	 & w/o T & 3.791 & 7.266 & 6.148 & 7.315 & 51.90 & 51.69 \\*
	 & EGGPlacer & 3.978 & 7.361 & 6.130 & 7.294 & 51.99 & 51.63 \\
\addlinespace[2pt]
	\multirow{4}{*}{ariane136} & w/o A\&T & 2.890 & 8.556 & 6.993 & 8.249 & 52.32 & 52.09 \\*
	 & w/o A & 3.374 & 8.428 & 6.226 & 7.424 & 51.76 & 51.46 \\*
	 & w/o T & 4.038 & 7.691 & 6.186 & 7.401 & 52.97 & 52.69 \\*
	 & EGGPlacer & 4.504 & 7.815 & 6.084 & 7.297 & 52.39 & 52.11 \\
\addlinespace[2pt]
	\multirow{4}{*}{ariane81} & w/o A\&T & 2.699 & 7.978 & 5.153 & 6.446 & 58.91 & 56.96 \\*
	 & w/o A & 3.197 & 8.567 & 5.350 & 6.725 & 56.47 & 54.84 \\*
	 & w/o T & 4.375 & 7.768 & 5.271 & 6.681 & 56.04 & 54.22 \\*
	 & EGGPlacer & 5.349 & 7.866 & 5.253 & 6.661 & 55.53 & 53.81 \\
\addlinespace[2pt]
	\multirow{4}{*}{bp} & w/o A\&T & 0.454 & 11.310 & 9.259 & 11.835 & 53.65 & 53.31 \\*
	 & w/o A & 0.539 & 11.081 & 8.491 & 9.883 & 54.39 & 53.82 \\*
	 & w/o T & 0.659 & 10.627 & 8.776 & 10.362 & 53.76 & 53.38 \\*
	 & EGGPlacer & 0.718 & 11.005 & 8.637 & 10.322 & 54.40 & 54.05 \\
\addlinespace[2pt]
	\multirow{4}{*}{bp68} & w/o A\&T & 0.684 & 27.919 & 15.333 & 17.622 & 70.42 & 69.26 \\*
	 & w/o A & 0.753 & 31.171 & 17.269 & 19.618 & 69.40 & 68.26 \\*
	 & w/o T & 1.204 & 19.112 & 11.355 & 12.686 & 71.26 & 69.59 \\*
	 & EGGPlacer & 1.416 & 20.513 & 11.425 & 12.968 & 69.45 & 68.28 \\
\addlinespace[2pt]
	\multirow{4}{*}{bp\_be} & w/o A\&T & 0.282 & 5.069 & 3.592 & 4.666 & 61.58 & 61.34 \\*
	 & w/o A & 0.290 & 5.004 & 3.402 & 4.401 & 61.47 & 61.16 \\*
	 & w/o T & 0.418 & 4.772 & 3.494 & 4.270 & 60.73 & 60.51 \\*
	 & EGGPlacer & 0.420 & 4.888 & 3.476 & 4.318 & 60.95 & 60.66 \\
\addlinespace[2pt]
	\multirow{4}{*}{bp\_be12} & w/o A\&T & 0.301 & 5.962 & 3.560 & 4.053 & 59.77 & 58.85 \\*
	 & w/o A & 0.342 & 6.080 & 3.604 & 4.227 & 59.76 & 58.92 \\*
	 & w/o T & 0.363 & 5.780 & 3.441 & 4.159 & 60.79 & 60.04 \\*
	 & EGGPlacer & 0.402 & 5.865 & 3.455 & 4.131 & 58.61 & 57.95 \\
\addlinespace[2pt]
	\multirow{4}{*}{bp\_fe} & w/o A\&T & 0.176 & 3.146 & 2.225 & 2.816 & 70.88 & 70.47 \\*
	 & w/o A & 0.181 & 3.217 & 2.358 & 3.213 & 70.99 & 70.72 \\*
	 & w/o T & 0.218 & 2.983 & 2.300 & 3.112 & 71.18 & 70.82 \\*
	 & EGGPlacer & 0.237 & 3.017 & 2.374 & 3.479 & 70.95 & 70.68 \\
\addlinespace[2pt]
	\multirow{4}{*}{bp\_fe38} & w/o A\&T & 0.976 & 4.573 & 3.182 & 3.607 & 64.34 & 63.42 \\*
	 & w/o A & 1.031 & 4.718 & 2.785 & 3.108 & 63.11 & 62.32 \\*
	 & w/o T & 1.548 & 4.602 & 2.673 & 2.984 & 64.08 & 63.20 \\*
	 & EGGPlacer & 1.747 & 4.682 & 2.719 & 3.002 & 64.44 & 63.68 \\
\addlinespace[2pt]
	\multirow{4}{*}{bp\_multi} & w/o A\&T & 0.451 & 6.245 & 4.904 & 5.622 & 64.04 & 63.69 \\*
	 & w/o A & 0.530 & 6.802 & 5.040 & 5.810 & 64.27 & 63.85 \\*
	 & w/o T & 0.729 & 5.940 & 4.876 & 5.589 & 63.38 & 63.05 \\*
	 & EGGPlacer & 0.713 & 5.834 & 4.918 & 5.653 & 63.87 & 63.48 \\
\addlinespace[2pt]
	\multirow{4}{*}{bp\_multi57} & w/o A\&T & 0.598 & 8.708 & 6.716 & 7.503 & 62.58 & 60.48 \\*
	 & w/o A & 0.842 & 9.242 & 6.723 & 7.466 & 58.69 & 57.54 \\*
	 & w/o T & 0.945 & 7.296 & 5.592 & 6.309 & 61.41 & 59.89 \\*
	 & EGGPlacer & 1.036 & 7.431 & 5.717 & 6.377 & 60.94 & 59.45 \\
\addlinespace[2pt]
	\multirow{4}{*}{dft68} & w/o A\&T & 2.955 & 2.224 & 1.161 & 1.306 & 73.73 & 73.21 \\*
	 & w/o A & 3.530 & 2.548 & 1.322 & 1.474 & 75.31 & 74.59 \\*
	 & w/o T & 3.533 & 2.068 & 1.087 & 1.233 & 73.89 & 73.49 \\*
	 & EGGPlacer & 4.454 & 2.328 & 1.212 & 1.362 & 74.85 & 74.08 \\
\addlinespace[2pt]
	\multirow{4}{*}{ethernet} & w/o A\&T & 0.311 & 1.429 & 0.861 & 1.092 & 54.42 & 54.28 \\*
	 & w/o A & 0.245 & 1.490 & 0.923 & 1.172 & 54.27 & 54.19 \\*
	 & w/o T & 0.166 & 1.381 & 0.877 & 1.109 & 55.90 & 55.72 \\*
	 & EGGPlacer & 0.185 & 1.387 & 0.888 & 1.127 & 55.87 & 55.63 \\
\addlinespace[2pt]
	\multirow{4}{*}{isa\_npu} & w/o A\&T & 0.469 & 11.192 & 6.300 & 7.789 & 52.06 & 51.30 \\*
	 & w/o A & 0.546 & 11.140 & 6.243 & 7.723 & 52.63 & 51.72 \\*
	 & w/o T & 0.755 & 11.350 & 6.289 & 7.778 & 51.72 & 50.97 \\*
	 & EGGPlacer & 0.958 & 11.617 & 6.327 & 7.815 & 51.62 & 51.17 \\
\addlinespace[2pt]
	\multirow{4}{*}{mor1kx} & w/o A\&T & 0.626 & 3.015 & 1.900 & 2.533 & 57.99 & 57.78 \\*
	 & w/o A & 0.857 & 3.050 & 1.910 & 2.547 & 59.32 & 58.83 \\*
	 & w/o T & 0.823 & 2.927 & 1.908 & 2.561 & 58.27 & 58.03 \\*
	 & EGGPlacer & 0.790 & 2.959 & 1.972 & 2.668 & 58.85 & 58.61 \\
\addlinespace[2pt]
	\multirow{4}{*}{or1200} & w/o A\&T & 0.698 & 1.589 & 0.851 & 1.029 & 77.56 & 76.82 \\*
	 & w/o A & 0.798 & 1.562 & 0.870 & 1.046 & 76.95 & 76.29 \\*
	 & w/o T & 0.973 & 1.385 & 0.801 & 0.943 & 77.43 & 76.85 \\*
	 & EGGPlacer & 1.017 & 1.430 & 0.797 & 0.941 & 75.73 & 75.06 \\
\addlinespace[2pt]
	\multirow{4}{*}{sw} & w/o A\&T & 1.355 & 5.312 & 4.556 & 5.391 & 56.35 & 56.03 \\*
	 & w/o A & 1.844 & 5.945 & 4.452 & 5.235 & 55.44 & 55.20 \\*
	 & w/o T & 2.089 & 4.978 & 4.093 & 4.869 & 56.41 & 55.97 \\*
	 & EGGPlacer & 2.238 & 5.275 & 4.277 & 5.046 & 56.74 & 56.38 \\
\addlinespace[2pt]
	\multirow{4}{*}{sw43} & w/o A\&T & 1.450 & 7.017 & 4.824 & 5.573 & 54.93 & 54.45 \\*
	 & w/o A & 1.656 & 6.664 & 4.404 & 5.249 & 54.11 & 53.59 \\*
	 & w/o T & 2.653 & 6.156 & 4.481 & 5.262 & 54.05 & 53.61 \\*
	 & EGGPlacer & 2.881 & 6.330 & 4.254 & 5.040 & 53.81 & 53.36 \\
\addlinespace[2pt]
	\multirow{4}{*}{vga\_lcd} & w/o A\&T & 1.094 & 2.501 & 1.866 & 2.705 & 72.47 & 71.81 \\*
	 & w/o A & 1.281 & 2.380 & 1.745 & 2.580 & 71.44 & 70.56 \\*
	 & w/o T & 1.306 & 2.057 & 1.513 & 2.510 & 73.18 & 72.39 \\*
	 & EGGPlacer & 1.183 & 2.281 & 1.645 & 2.593 & 73.31 & 72.22 \\
\addlinespace[2pt]
	\multirow{4}{*}{VeriGPU} & w/o A\&T & 0.029 & 1.872 & 1.133 & 1.621 & 67.95 & 67.19 \\*
	 & w/o A & 0.034 & 1.892 & 1.180 & 1.666 & 62.77 & 62.38 \\*
	 & w/o T & 0.046 & 1.801 & 1.116 & 1.597 & 67.14 & 66.31 \\*
	 & EGGPlacer & 0.045 & 1.804 & 1.134 & 1.609 & 65.60 & 64.89 \\
\end{longtable}
}


\clearpage

\end{document}